\documentclass{elsart}
\usepackage{graphicx}

\usepackage{color}

\newcommand{\vev}[1]{\left\langle#1\right\rangle}

\newcommand{\jtem}[2]{\par\hangafter0\hangindent#1
               \noindent\llap{#2\enspace}\ignorespaces}
\newcommand{\piz}[0]{{\pi^0}}               

\newcommand{\simge}
   {\mathrel{\rlap{\raise 0.511ex
       \hbox{$>$}}{\lower 0.511ex \hbox{$\sim$}}}}

\newcommand{\simle}
     {\mathrel{\rlap{\raise 0.511ex
      \hbox{$<$}}{\lower 0.511ex \hbox{$\sim$}}}}
            
\font\tenmib=cmmib10 
\newcommand{\bold}[1]{\hbox{\tenmib #1}}
\newcommand{\mbf}[1]{\hbox{\boldmath $#1$}}

\begin{document} 

\begin{frontmatter}
\title{Energy flow in a hadronic cascade: Application to hadron
calorimetry}

\author{Donald E. Groom} 
\corauth{Tel: 1 510 486 6788; fax: +1 510 486 4799}
\ead{deg@lbl.gov}
\address{Lawrence Berkeley National Laboratory, 50R6008,
Berkeley, CA 94720, USA}

\begin{abstract}           

The hadronic cascade description developed in an earlier paper is extended
to the response of an idealized fine-sampling hadron calorimeter.
Calorimeter response is largely determined by the transfer of energy $E_\piz $
from the hadronic to the electromagnetic sector via $\pi^0$ production.
Fluctuations in this quantity produce the ``constant term" in hadron
calorimeter resolution. The increase of its fractional mean, $f_{\pi^0}^0
= \vev{E_\piz }/E$, with increasing incident energy $E$ causes the energy
dependence of the $\pi/e$ ratio in a noncompensating calorimeter. The mean
hadronic energy fraction, $f^0_h = 1-f_{\pi^0}^0$, was shown to scale
very nearly as a power law in $E$: $f^0_h = (E/E_0)^{m-1}$, where
$E_0\approx1$~GeV for pions, and $m\approx0.83$. It follows that
$\pi/e=1-(1-h/e)(E/E_0)^{m-1}$, where electromagnetic and hadronic energy
deposits are detected with efficiencies $e$ and $h$, respectively.
If the mean fraction of $f^0_h$ which is deposited as nuclear 
gamma rays is $f^0_\gamma$, then the expression becomes 
$\pi/e=1-(1-h^\prime/e)(1-f^0_\gamma)(E/E_0)^{m-1}$.
Fluctuations in these quantities, along with sampling fluctuations, are
incorporated to give an overall understanding of resolution, which is
different from the usual treatments in interesting ways. The conceptual
framework is also extended to the response to jets and the difference     
between $\pi$ and $p$ response.

\end{abstract}
\end{frontmatter}

\section{Introduction}

In Paper~I\cite{gabriel94} we developed a conceptual basis for 
understanding the division between hadronic and electromagnetic (actually $\pi^0$)
energy 
deposition in a contained hadronic cascade.\footnote{Most of the content 
of Paper~I was first presented at the 1989 Workshop on Calorimetry for the 
Superconducting Super Collider\cite{tuscaloosa90}.} The model 
``calorimeter'' was a very large iron or lead cylinder, with no energy 
leakage except via muons, neutrinos, and front-surface albedo losses. 
Extensive Monte Carlo simulations gave results in good agreement with 
test-beam measurements. The relevant conclusions of the paper were that:

\begin{enumerate}

\item All significant hadronic energy deposition is by low-energy 
particles\break 
($\simle1$~GeV), whose energy and species distribution in a 
given medium is 
independent of the energy or species of the incident hadron. (Hadronic 
energy was defined as all energy not carried away by $\pi^0$ decay 
photons.)  The existence of this ``universal low-energy 
hadron (and nuclear gamma ray) spectrum" makes it possible to define an 
energy-independent 
efficiency $h$ for the conversion of this energy into a visible signal in 
a fine-sampling calorimeter.

\item In each high-energy collision of the hadronic cascade, a significant 
fraction (typically 1/4) of the energy is lost from further hadronic activity 
via $\pi^0$ production.  A sequence of 
high-energy hadronic collisions bleeds off a larger and larger fraction 
of the energy as the incident energy $E$ increases.  The net fraction 
transferred to the $\pi^0$ sector in a given cascade is $f_{\pi^0}$, and the 
mean $\pi^0$ fraction is $f_{\pi^0}^0$.%
\footnote{Throughout this paper, the superscript 0 indicates the mean of a stochastic
variable, e.g.~$f_\piz^0 = \vev{f_\piz}$. In most of the literature the quantity without the superscript
indicates the mean.}
 This one-way flow is illustrated in 
Fig.~\ref{fig:EFlowSimple}.\footnote{Wigmans points out that the actual number of $\pi^0$'s 
produced is quite small\cite{wigmansbook}.}

\begin{figure}
\centerline{\includegraphics{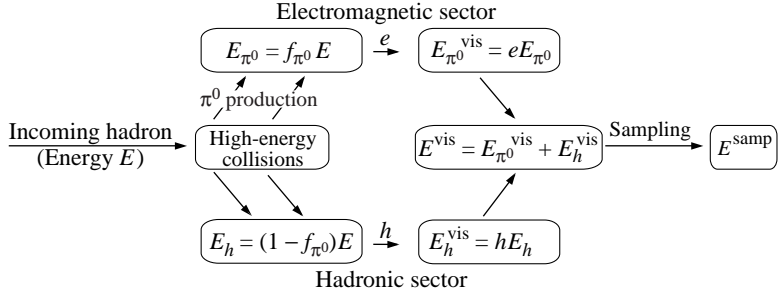}}
\caption{Energy flow in a hadronic cascade. A fraction $f_{\pi^0}$ (with 
energy-dependent mean $f_{\pi^0}^0$) is transferred to the 
electromagnetic sector through $\pi^0$ production in repeated hadronic 
inelastic collisions.  The $\pi^0$ and hadronic energy deposits after the 
division are separately stochastic, and so must be treated as parallel statistical
processes.  Each produces a visible signal, whose sum $E^{\rm vis}$ is 
sampled.
}
\label{fig:EFlowSimple}
\end{figure}

\item 
In particular, the mean fraction of the energy $f^0_h=1-f^0_{\pi^0}$ in 
the hadronic sector scales very nearly as a power of the incident energy, 
\begin{equation} 
f^0_h (E) = (E/E_0)^{m-1}\ , 
\label{eqn:powerlaw}
\end{equation} 
where $m\approx0.83$ (with some mild absorber $Z$ dependence) and 
$E_0\approx1$~GeV for pions and $\approx 2.6$~GeV for protons (again, with
some $Z$ dependence). 
Physically, $m$ is related to the mean number of secondaries and the mean 
energy fraction going into $\pi^0$'s in any given collision in the 
cascade, and $E_0$ is the energy at which multiple pion production becomes 
significant.  Both must be determined by experiment for a given calorimeter.

\item 
It was predicted that a calorimeter would have a different response 
to a proton than to a pion.

\end{enumerate}

The observations pertain equally well to a homogeneous or 
fine-sampling calorimeter, and have
significant implications for its response and resolution.   ``Fine-sampling"
means that absorber and sensor elements are thin compared to both the
em radiation length and the neutron interaction length.  
It has the same structure throughout:
no separate front em compartment or rear catcher. 
It can be an inorganic crystal calorimeter,
a uranium/liquid argon calorimeter, or a lead/scintillator-fiber calorimeter.

The power-law {\it approximation} given in Eq.~(\ref{eqn:powerlaw})
is just that, for reasons discussed in  Paper~I. It seems to work well over the 
energy range of available test-beam data, about 10~GeV to 375~GeV, 
and it has the required asymptotic properties: 
It is everywhere positive, and $f^0_h \to 0$ ($\pi/e\to1$) as $E\to\infty$.  The physical 
assumptions it is based upon become less dependable at very high energies 
and are not valid  at energies below the  threshold for multiple pion 
production.

As far as possible, results in this paper are obtained without recourse 
to the power-law approximation for $f^0_h$, in order to obtain more general 
results than those relying on this more approximate form.

\begin{figure}
\centerline{\includegraphics{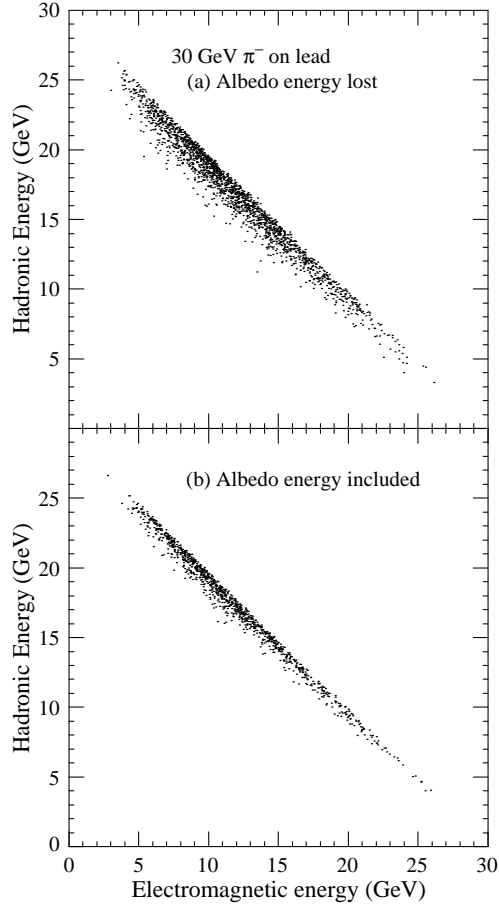}}
\caption{Calculated energy deposit  distributions for 30 GeV negative pions incident on a lead 
``calorimeter.''   In case (a) backscattered energy is lost; in (b) it is retained.
}
\label{fig:comp30_bothdot}
\end{figure}

Most of the results reported in this paper can be found in 
the {\it Proceedings} of various conferences and 
workshops\cite{tuscaloosa90,barcelona89,snowmass90,fortworth90,aachen90,capri91,tucson97}.
The Monte Carlo results used in these papers are often based on now-superseded 
versions of hadronic cascade simulation codes\cite{DPM}, the oldest being 
FLUKA86. In particular, nuclear gamma rays were not included, so that the em
deposit is exclusively via $\pi^0$ production.
Since these versions many improvements in the codes have been made, 
e.g., improvements in FLUKA by Ferrari and 
Sala\cite{aliceFLUKA}, especially in the nuclear physics modeling.  
The failings of the old code are apparent in 
Fig.~\ref{fig:comp30_bothdot}(b), for example, where the points fall below 
the 45$^\circ$ line because of unscored hadronic energy. A large fraction of
the unscored energy is evidently that of nuclear gamma rays. On the other 
hand, $\piz$ energy deposition was very well described\cite{EGS4} 
and can be trusted.  In Paper~I we reported simulations with MARS10, HETC, 
and FLUKA, which, though based on different high-energy interaction 
models, were in excellent agreement.  Since in this paper I depend {\em 
only} upon the high-energy division between the $\pi^0$ and hadronic sectors, 
calculations based on the older code have not been repeated.

In Sec.~\ref{sec:pi_over_e}, I distinguish
between em energy deposit by $\pi^0$ decay photons and by nuclear de-excitation gamma
rays.  A fraction $f_{\piz}$ of the energy is deposited via  $\piz$
decay, and a fraction $f_h f_\gamma$ by nuclear gamma rays {\it within the 
acceptance gate}, where $f_h =1-f_\piz$.  The total em 
deposit  is $E_\piz  + E_\gamma = E(f_\piz  + f_h f_\gamma)$. 

Recent developments are incorporated, some of which were predicted or 
discussed in Paper I. These include Cherenkov readout\cite{qcal97},
which is sensitive only to em ($\pi^0$ and nuclear gamma)  energy deposition,
and observation of the $\pi/p$ response difference\cite{wigmans_pi_p_98}.

Central to the paper is the discussion of resolution, where 
conditional probability distribution functions
(p.d.f.'s) are combined to account for parallel, independent stochastic processes.

Hadron calorimetry is a well-traveled road, explored in hundreds, if not 
thousands, of papers over several decades. The object here 
is to present a broad-brush treatment of 
hadronic cascades in a simplified generic calorimeter, in hopes 
that a somewhat nonstandard approach can contribute to our physical  
understanding of a real calorimeter.  Real 
calorimeters, with front em compartments, rear catchers, leakage, crack 
corrections, jet finding algorithms, and a myriad of other features, are 
described in dozens of test-beam 
study results, as well as in published studies of compensation, the role 
of neutrons, and other matters.  These are discussed in detail in Wigmans' 
book\cite{wigmansbook} and review\cite{wigmansAnnRev}, 
the review by Leroy and Rancoita\cite{leroy00},  and in their many citations.  
None of these practical problems are discussed here.

\section{Albedo and  \mbf{f_{\pi^0}}\label{sec:albedo}}

The $\pi^0$ fraction $f_{\pi^0}= E_{\pi^0}/E$ increases with energy, but at any given 
energy it is subject to large fluctuations. FLUKA 
simulations of the $\pi^0$/hadronic energy division are shown in 
Fig.~\ref{fig:comp30_bothdot}. The model absorber consisted of a large 
lead cylinder (50 cm radius, 250 cm long) in which the first 25~cm 
(about 1.5 interaction lengths) was treated as 
a separate region. In Fig~\ref{fig:comp30_bothdot}(a) no distinction is 
made between the regions, while in ~\ref{fig:comp30_bothdot}(b)  interaction of the 
incident pion was not permitted in the front section, but energy deposited
there is included.  It acted as a catcher for back-scattered interaction debris.
The distribution about the ideal $E_h=1-E_{\pi^0}$ 
shows less scatter in \ref{fig:comp30_bothdot}(b) because front-face, 
or albedo, losses are included. Most albedo loss comes from 
backward or backscattered products of the first collision; when the first 
interaction occurs deep in the detector there is essentially no albedo loss. 
Runs at 50 GeV with and without an ``albedo catcher'' show an average 
difference in deposited energy is 0.43~GeV, or 0.8\%. Out of 1000 cascades 
50\% lost less than 0.2~GeV, and 3.4\% lost more than 2~GeV.  In the 
simulations the amount of lost albedo energy rises only slowly with 
increasing incident energy, as might be expected. While these losses are 
not totally negligible, I omit them from resolution considerations in Sec,~\ref{sec:resolution} 
because (a) the distribution is sharply peaked at near-zero loss, and
(b) the losses are small, particularly at higher energies.

\begin{figure}
\centerline{\includegraphics{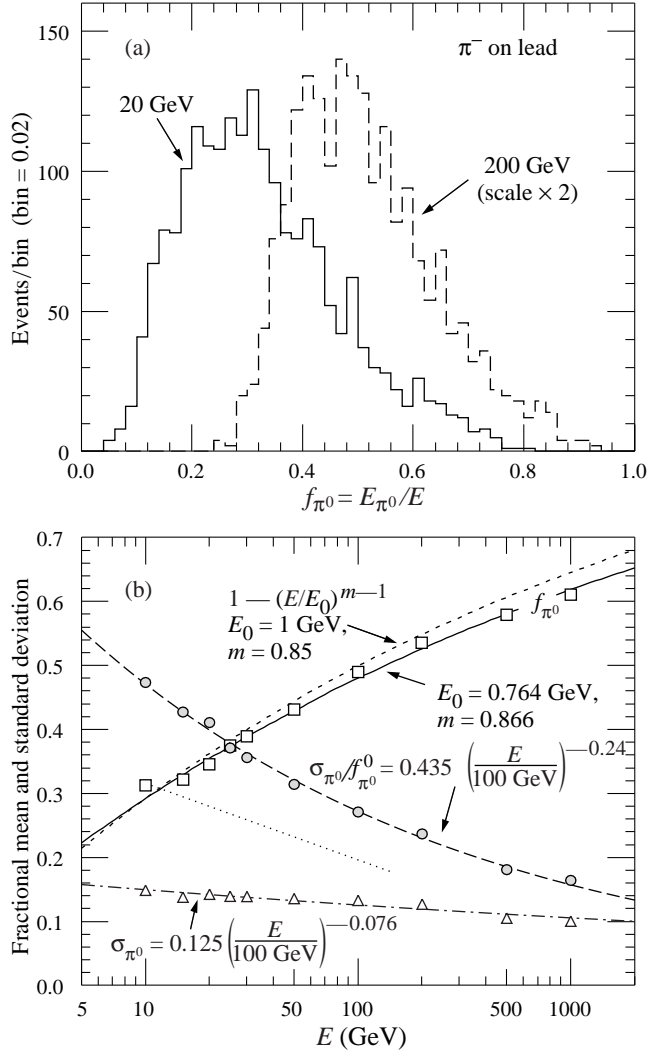}}
\caption{FLUKA simulations for negative pions  incident on a lead ``calorimeter.''
  (a) Distribution of the $\piz$ energy deposit for 20 and 200 GeV incident pions,
 and (b) energy dependence of the mean ($\vev{f_\piz}=f^0_\piz$),  standard deviation 
  ($\sigma_\piz$) of the $f_\piz$ 
 distribution, and standard deviation relative to the mean ($\sigma_\piz/f_0$).  
 The dotted line is discussed in the text.
}
\label{fig:f0_distr}
\end{figure}

For reasons discussed in the introduction, the points shown in Fig.~\ref{fig:comp30_bothdot}
scatter below the 
45$^\circ$ line because older versions of FLUKA did not account for all of 
the hadronic energy deposit, even in the absence of albedo losses.
Presumably most of this downward scatter (about 15\% in the worst case) is the result
of the program's failure to tally nuclear gamma rays, most of which come from 
de-excitation following slow neutron
capture by nuclei. According to Ferrari and Sala\cite{ferrari00}, these might account for 
nearly 10\% (Fe) or 20\% (Pb) of the $\pi^0+\gamma$ fraction, or 5\%--10\% of the total
energy deposit.   These contributions scale with the hadronic fraction, not the $\pi^0$ fraction.
Given that the hadronic fraction is underestimated by this fraction in this simulation, 
it is better to take the hadronic fraction as
\begin{equation}
f_h \equiv 1- f_{\pi^0},\ {\rm{or}}\  E_h \equiv E-E_{\pi^0} \ .
\label{eqn:Eh_restored}
\end{equation}

As the number of Monte Carlo events in the sample increases, the $(E_h,  E_{\pi^0})$ 
distribution projected onto the $ E_{\pi^0}$ axis approaches the 
marginal distribution $\Pi( f_{\pi^0})$, the p.d.f\  of the $\pi^0$ energy fraction. 
Two  (unnormalized) FLUKA-generated examples of 
$f_\piz \equiv E_{\pi^0}/E$ distributions are shown in Fig.~\ref{fig:f0_distr}(a). The mean 
and standard deviations are shown in Fig.~\ref{fig:f0_distr}(b). 

The fractional mean $f_\piz^0$ moves slowly to the right with increasing energy, 
and can be represented by $f_{\pi^0}^0=1-(E/E_0)^{m-1}$.  As it does so, the rms width
of the distribution decreases only slowly (presumably because of increasing $\piz$ multiplicity), 
and  is well-represented by 
\footnote{A slightly better fit is obtained with $0.126-0.0099\ln(E/100\, \rm{GeV})$.} 
\begin{equation}
\sigma_\piz=12.5\%\times(E/100 \,\rm{GeV})^{-0.076} \ .%
\label{eqn:pwrlaw_sigmapiz}
\end{equation}There is no physical basis for this functional form, except that
it remains positive as $E\to\infty$. As the incident energy becomes very large, the 
distribution ``crowds'' the right limit, and the variance should approach zero.  
The fits to real data (Fig.~\ref{fig:groom_final_resplot} and Tbl.~\ref{tab:fitparameters}) 
yield somewhat different values for the multiplier and exponent, which in any case should
vary from case to case.

Alternatively, one might have chosen to express the width as a fraction of the mean rather 
than as a fraction of the total incident energy, or as $\sigma_\piz/f_\piz^0$ rather than
$\sigma_\piz$.  A power-law fit to the Monte Carlo data in this form is indicated by the 
dashed line in Fig.~\ref{fig:f0_distr}(b). The strong energy dependence just reflects the
energy dependence of $f_\piz^0$, and the mildness of the energy dependence 
of $\sigma_\piz$ is obscured.  

The dotted line in Fig.~\ref{fig:f0_distr}(b) is a fit by Acosta et~al. 
to  SPACAL data for $\sigma_{\rm rms}(f_1)/f_1$ in the range 10--150\,GeV, where
 ``1'' refers to the central tower\cite{spacal92a}.   The authors show that
the $f_1$ distribution is a good representation of the $f_\piz$ distribution.  The fit is given
as $0.435-0.052\ln(E)$.  SPACAL was a lead/scin\-til\-lator fiber calorimeter, while our model
is a solid lead cylinder.  The test beam events contained contributions from nuclear gamma 
rays.  Even so, the difference between the dotted line and the FLUKA-simulation dashed line is
difficult to understand. 

The dimensionless 
``coefficient of skewness," $\gamma_1= \mu_3/\sigma_\piz^3$ (where 
$\mu_3$ is the third moment about the mean), is constant to within the 
Monte Carlo statistics with a value near 0.6. There are no significant 
higher moments within the sensitivity of the simulations.  One might expect to see 
the skewness change sign; the tail should move from the right to the left side of the most 
probable value at very high energies. This transition has not yet been observed.

It is instructive to examine a continuous distribution with similar properties:
the Beta distribution $f(x;p,q) = x^{q-1}(1-x)^{s-1}/B(p,q)$, where the normalizing 
constant $B(p,q)$ is the Beta function\cite{betafunction}. 
It may be thought of as ``a continuous version of the binomial distribution:" 
It is defined only for $0 \leq x \leq 1$, is zero at both 
limits, and the  peak position, variance, and other properties depend on
 $p$ and $q$. For $p>2$ and 
$q>2$ it has zero derivative at both limits. If the mean is less than 0.5 the distribution is skewed 
to the right, as is the case for $\Pi(f_{\pi^0})$; for larger means the distribution is skewed to the left.
The skewness of $\Pi(f_{\pi^0})$ remains positive for means much greater than 0.5, but,
as explained above, might be expected to change sign as the mean approaches unity.

Different or better cascade simulations would be expected to produce distributions with 
somewhat different shapes and different moments.  What is of consequence here
is that a function $\Pi(E_{\pi^0})$ exists which describes the energy distribution of the
$\pi^0$'s for a given primary energy $E$; no significant conclusions in this paper depend
upon the details.

\begin{figure}
\label{fig:compf30_corr_scat}
\centerline{\includegraphics{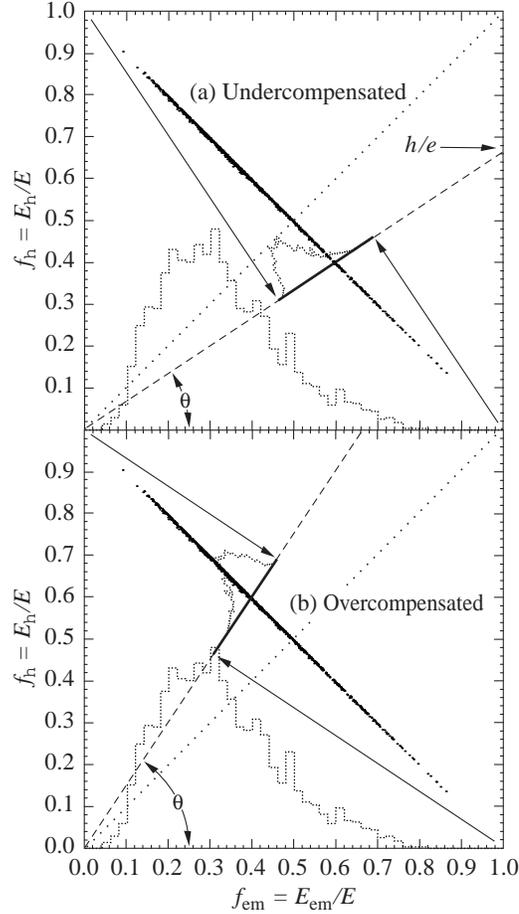}}
\caption{The projection of ``events'' onto the diagonal line at $\theta = 
\tan^{-1}(h/e)$ gives the energy distribution for a calorimeter in the 
absence of any other fluctuations.  Events must fall along the thick solid 
segments.  One can imagine the projection as $\theta$ increases from 
0$^\circ$ ($h/e=0$, large dotted histogram through the undercompensated case shown in 
(a)) to 45$^\circ$ ($h/e=1$, full 
compensation) where it approaches a $\delta$-function.  For $\theta>45^\circ$ 
($h/e>1$, shown in (b)), the 
calorimeter is overcompensated,  and the skewness changes sign.
}
\end{figure}

The contribution of $\Pi(f_\piz)$
to the calorimeter resolution can be understood by a geometrical construction. 
Figure~\ref{fig:compf30_corr_scat} shows the same MC ``events'' as 
Fig.~\ref{fig:comp30_bothdot}(b), but with the lost hadronic energy 
restored as per Eq.~(\ref{eqn:Eh_restored}) (except for some vertical 
scatter retained for clarity).  The observed energy distribution in the 
absence of sampling fluctuations is the projection of this distribution 
onto a diagonal line at $\theta = \tan^{-1}(h/e)$.  The limits of the projected
distribution ($f_{\pi^0}=0$ and $f_{\pi^0}=1$) are shown by the arrows. All of the events 
thus project onto the solid segment  of the line, with length 
$|\cos\theta-\sin\theta|= (1-h/e)\cos\theta$.  The sampled $\Pi(f_\piz)$ distribution 
in Fig.~\ref{fig:f0_distr}(b) is replotted as the
histogram along the $f_\piz$ axis.  A point at $f_\piz=1$ 
projects to the end of the solid segment, so the energy scale along this 
axis is foreshortened by $\cos\theta$.  The length of the solid line segment, rescale 
by $1/\cos\theta$, is $(1-h/e)$.  The  fractional 
standard deviation of $\Pi(f_\piz)$, $\sigma_\piz$, also scales as 
$(1-h/e)$.  It thus contributes $(1-h/e)\sigma_\piz$ (in quadrature) to the calorimeter 
resolution.

For $h/e=1$, $\theta=45^\circ$ and the distribution becomes a $\delta$-function. For 
$h/e>1$ (overcompensation) the distribution ``flips," with the tail on the low-energy side,
since the $\piz$-rich events in the high-energy tail of $\Pi(f_\piz)$ now contribute less energy to the
cascade than do the hadrons.   

Experimental verification of this situation is at least strongly suggested by the WA~78 results obtained 
with a uranium-scintillator plate calorimeter\cite{wa78}. The bulk of the 
energy was deposited in the upstream
``Section A," which in some configurations was Fe-scintillator and in others U-scintillator.  
Unweighted energy distributions are shown in the paper's Fig.~3 for the Fe-scintillator
case (Fe25) and in Fig.~8 for the most uniform U-scintillator case (U15). While other resolution 
effects broaden the distributions, the distributions are nonetheless skewed to the right for Fe25
and skewed to the left for U15.  The dotplots in their Figs.~4 and 9 show uncorrected $E_{\rm tot}$
vs $A_{\rm max}$, the maximum energy deposited in one of the scintillator sheets. Large deposits
indicate large em shower activity, and hence $\piz$-rich events. The mean of the distribution slopes
upward in the Fe25 case and downward in the U15 case, again providing evidence for  
the inversion of the distribution.

\section{$\mbf{\pi}$/\bold{e}\label{sec:pi_over_e}}

An electromagnetic shower initiated by an electron or $\pi^0$-decay photons produces 
a visible signal  (potentially observable via ionization or Cherenkov light) 
in a calorimeter with efficiency 
$e$.  Most of the ionization is by  electrons and positrons with energies below
the critical energy, of  order 10 MeV 
(21.8 MeV for iron, 7.0~MeV for uranium).  The response, here temporarily called ``$e$,'' is usually 
linear in the incident energy $E$, and so serves to calibrate the energy scale:
\begin{equation}
\hbox{``}e\hbox{''}  = e\,E
\label{eqn:e_response}
\end{equation}

As shown in Paper~I, the visible signal produced by hadron interactions also comes predominately 
from low-energy ionizing particles whose spectra and relative abundance are independent of the 
incident hadron energy.  Many mechanisms are at play, including endothermic nuclear spallation.  Neutrons play an especially 
significant role\cite{wigmans98_neut}.  These mechanisms are exhaustively treated
in the literature; for example, in 
Refs.~\cite{wigmansbook,wigmansAnnRev,leroy00,ferrari00}.
The sum of all the hadronic energy deposit mechanisms (excluding 
showers by $\pi^0$ decay photons) 
produces an observable signal with efficiency $h$.  In most cases 
$h/e\leq1$.  For a mean hadronic fraction $f^0_h = 1 - f_{\pi^0}^0$,
\begin{eqnarray}
\hbox{``}\pi\hbox{''}  =  &e\,f_{\pi^0}^0E+h\,f^0_h E
\nonumber\\
= & eE[1-(1-h/e)f^0_h]\ .
\label{eqn:meanvis}
\end{eqnarray}
In the case of an an incident pion, the response relative to an electron is
\begin{equation}
\pi/e= 1-(1-h/e)f^0_h \ .
\label{eqn:pioverE_hadfrac}
\end{equation}
Specializing to our power-law form for $f^0_h$,
\begin{equation}
\pi/e= 1-(1-h/e)(E/E_0)^{m-1} \equiv 1-aE^{m-1}\ ,
\label{eqn:pioverE}
\end{equation}
where, as above, $m\approx0.82$ to 0.86.  ($\pi/e$ is only defined for an 
ensemble of events, so it is implicitly a mean value.)  Since the physics 
leading to the power law involves a multistep cascade, it is not expected 
to be dependable below 5--10~GeV.  Only $m$ and $a=(1-h/e)E_0^{1-m}$ can be 
obtained from fits to data, at least in a single-readout calorimeter. 
For incident pions  (not protons) a range of $E_0$'s 
near 0.7--1.0~GeV fit almost as well because $E_0$ is raised to a small power. 
{\em The ratio h/e cannot be obtained from a 
measurement of $\pi/e$ as a function of energy without other information 
or some assumption about} $E_0$.

\begin{figure}
\centerline{\includegraphics{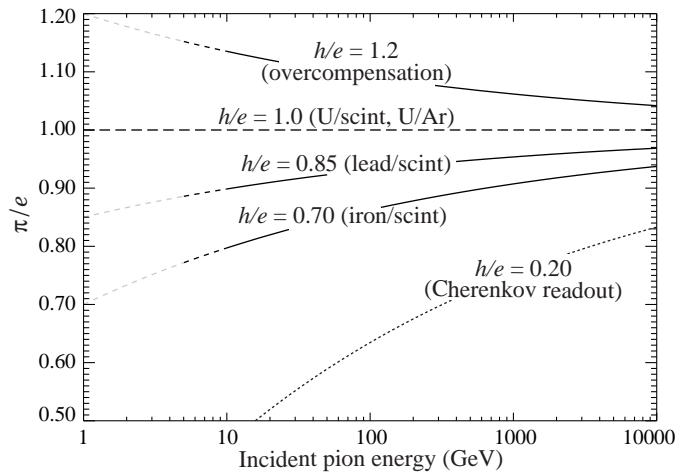}}
\caption{Energy dependence of  $\pi/e$ expected for several 
values of $h/e$ with the assumption that $E_0 = 1$~GeV. For almost all 
calorimeters, $h/e<1$.  The value for a given combination (U/scint, etc.)
depends on the actual configuration. The lower dotted 
line, for $h/e\ll1$, should be applicable to a calorimeter with Cherenkov 
readout.  The power-law description is not expected to 
be dependable below about 10~GeV, but nonetheless seems adequate at 
5~GeV.
}
\label{fig:epi_theory}
\end{figure}
\begin{figure}
\centerline{\includegraphics{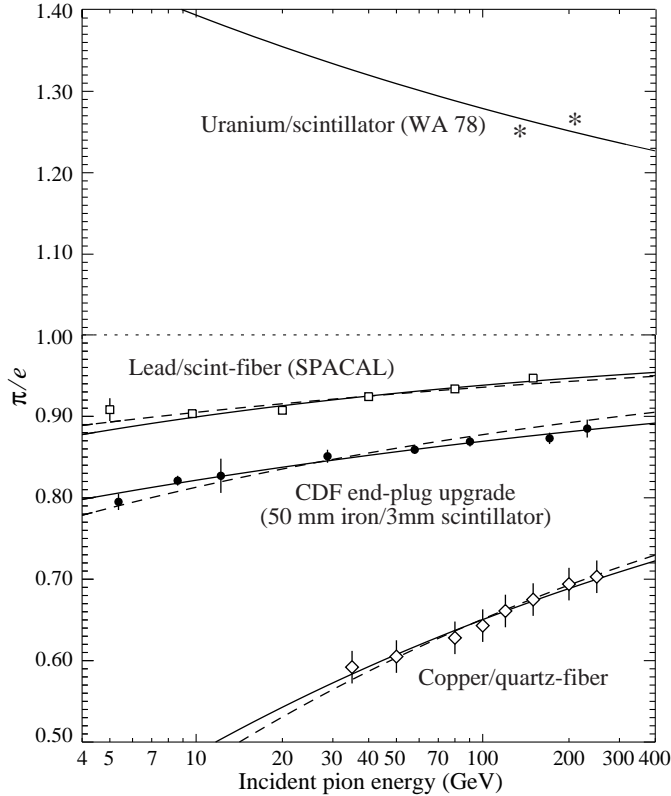}}
\caption{Fits to test-beam results for a lead/scintillator-fiber)\protect\cite{spacal91}, for
the CDF upgrade end-cap hadron calorimeter (50~mm iron/3~mm scintillator 
sheets)\protect\cite{jinbo98}, for a copper/quartz-fiber test 
calorimeter\protect\cite{qcal97}, and for the WA~78 uranium/scintillator 
calorimeter\cite{wa78}.   Fit parameters are given in Table~\ref{tab:pioe}.
}
\label{fig:datafits_pie}
\end{figure}

\begin{table}
\caption{Power law fits to a variety of $\pi/e$ measurements.
The ZEUS uranium/scintillator\protect\cite{behrens90} and D0
U/LAr\protect\cite{D0} calorimeters are so close to
compensating as to be uninteresting in this context.\label{tab:pioe}}
\begin{center}
\begin{tabular}{ccc ccc}
\hline
\hline
Calorimeter  & $m$ & $a$ & $\chi^2$ & $h/e$*
     & Expected $h/e$\\
\hline
SPACAL\cite{spacal91}&0.788
       & 0.164 &  9.2 & 0.836 & 0.853--0.895$^\dagger$ \\
(lead/scint-fiber)
       &{\em 0.830\,$^\ddagger$} & 0.141 & 14.0  & 0.859
       & 0.853--0.895$^\dagger$ \\
\hline
CDF end-plug had cal\cite{jinbo98}
        &0.865 & 0.244 & 2.7 & 0.756 & 0.667$^\dagger$ \\
(50 mm Fe/3 mm scint)
        &{\em 0.816\,$^\ddagger$} & 0.286 & 14.1 & 0.714 & 0.667$^\dagger$ \\
\hline
Copper/quartz-fiber\cite{qcal97}
                      & 0.833 & 0.753 & 2.6 & 0.247$^\S$ &\\
                    (QFCAL)& {\em 0.816\,$^\ddagger$} & 0.814 & 3.8 & 0.238&  \\
\hline
U/scint  (WA 78)\cite{wa78}
                      & 0.85$^\sharp$ & $-$0.555$^\sharp$ & -- & 1.555$^\sharp$ &\\
\hline
\hline 
\multicolumn{6}{l}{\footnotesize *Assuming $E_0=1$ GeV.}\\
\multicolumn{6}{l}{\footnotesize $\dagger$ Paper~I, Table 1. (The calorimeters
have only approximately the same structure.)}\\
\multicolumn{6}{l}{
\begin{minipage}{5.2in}
{\footnotesize$\ddagger\ $Dashed curves in Fig.~\ref{fig:datafits_pie}: 
$m$ held fixed at the value
given by the fitted line in Paper~I, Fig.~12. Error in $m$ from
this work is $\pm0.01$ to $\pm0.015$.} 
\end{minipage}}\\
\noalign{\vskip3pt}
\multicolumn{6}{l}{
\begin{minipage}{5.2in}
{\footnotesize $\S$ Akchurin et al.\ report $e/h\approx 5$\cite{qcal97}.
Virtually all of the hadronic Cherenkov signal can be accounted for as coming from relativistic
pions\cite{PbWO4}.}
\end{minipage}}\\
\noalign{\vskip3pt}
\multicolumn{6}{l}{
\begin{minipage}{5.2in}
{\footnotesize $\sharp$ Ref.~\cite{wa78} gives two data points, shown in 
Fig.~\ref{fig:datafits_pie}.  Here I assume $m=0.85$ and adjust $a$ for a best fit.}
\end{minipage}}\\
\end{tabular}
\end{center}
\vskip-0.1in
\end{table}

I emphasize again that the power-law representation is not empirical, but follows 
from an induction argument.  It has the correct asymptotic limit, since 
$f_{\pi^0}^0 \to 1$ as $E \to \infty$. For a $10^{19}$~eV proton-induced 
air shower, for example, $f_{\pi^0}^0 \approx 0.98$, in accord 
with the usual cosmic ray expectation and observation that nearly all the 
energy deposit at very high energies is electromagnetic.\footnote{This is 
an illustrative example only, because there is no expectation that $m$ will 
remain even relatively constant
over such a large energy range.  In addition, as much as  10\% of the 
energy is carried by muons and neutrinos from meson decay\cite{eloss_to_muons}.} 
The expected behavior of $\pi/e$ is shown in Fig.~\ref{fig:epi_theory}.

Representative fits of test-beam results to Eq.~(\ref{eqn:pioverE}) are 
shown in Fig.~\ref{fig:datafits_pie}. Solid curved are least-squares fits 
with both $m$ and $a$ allowed to vary, while dashed curves are fits
with $m$ is constrained to its nominal value from Paper~I.
Given the wide range of experimental data which have been fitted to test the
power law, one suspects that occasional disparate 
results (e.g., the low value of $m$ for the CDF end-plug calorimeter) 
indicate data reduction problems. 

Although the energy fraction carried by the nuclear gamma rays scales as
the hadronic fraction, it is detected with  efficiency $e$ (nearly).%
\footnote{The Compton electrons are low-energy to begin with, and a significant fraction of
their energy deposit occurs by ionization after they drop below Cherenkov threshold.  They are thus
detected with lower efficiency than are high-energy electromagnetic cascades. 
This nicety is ignored in the present discussion.}
Let $f_\gamma$ be the fraction of the hadronic energy deposited by 
nuclear gamma rays {\it within the electronic gate time}.  Its mean,  $f^0_\gamma$, is independent of both incident hadron energy 
and species via the ``universal spectrum" concept developed in Paper~I.
Thus $f_h\,f_\gamma$ of the incident energy is detected with 
efficiency $e$, and the total em fraction $f_{\rm em}$ is $f_\piz + f_h\,f_\gamma$.
The remaining $f_h\,(1-f_\gamma)$ is detected with redefined hadronic detection
efficiency $h^\prime$.  Then
\begin{eqnarray}
\pi/e =& (f^0_{\pi^0} + f^0_h f^0_\gamma) + (h^\prime/e)f^0_h (1-f^0_\gamma)
\nonumber\\
=& 1- \left(1-h^\prime/e)(1-f^0_\gamma\right) f^0_h
\nonumber\\
\approx&1-\left(1-h^\prime/e)(1-f^0_\gamma\right)(E/E_0)^{m-1}
\nonumber\\
\equiv& 1-a\,E^{m-1}\ ,
\label{eqn:pwrlawwithfgamma}
\end{eqnarray}
where, as elsewhere, the superscript zero indicates the mean.

The important point here is that the power-law description given by Eq.~\ref{eqn:pioverE} is
recovered, {\em even though part of the electromagnetic signal tracks with the hadronic
sector.}  It is immaterial whether we use $(1-h/e)$ or $(1-h^\prime/e)(1-f^0_\gamma)$, except that
we should remember that $h$ contains a nuclear gamma-ray component.

Energy deposit by nuclear gamma rays can account for 10\%--20\% of the total
energy deposit in materials such
as iron, and it can be even higher in high-$Z$ materials\cite{ferrari00}.
Since most of the nuclear  de-excitations are the result of slow neutron capture, nearly all 
of the gamma rays are emitted on
a time scale of hundreds of~ns (see Fig.~3.22 in Ref.~\cite{wigmansbook}).  
Electronic gate widths 
for calorimeter signals are made as short as possible, given the arrival time
spread of information in such large devices, and this turns out to be close to 100~ns.  
This means that most of the nuclear gamma energy
deposit happens outside of the sampling time. There are a few gammas from faster neutron
captures, but by and large most of the gamma signal is lost. Thus $f_\gamma$ as 
defined above is in the range of a few percent.  While the above development is of 
interest in showing that the power-law scaling does not need modification, it  is of not much
practical importance.

\section{\mbf{\pi}/\bold{p}}

We observed in  Paper~I that $f_{\pi^0}^0$ is 
larger for an incident charged pion than for an incident proton (or 
neutron).  This is a consequence of the fact that a leading hadron, 
carrying a large fraction of the energy, is likely to have the same quark 
number as the incident hadron.  If the collision is instigated by a 
charged pion, there is high probability that the leading hadron is a $\pi^0$, but
for an incident proton or neutron the leading hadron is most likely a baryon.

\begin{figure}
\centerline{\includegraphics{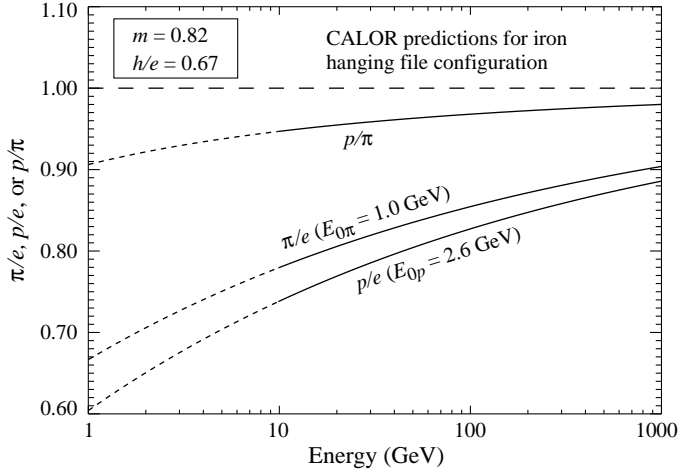}}
\caption{Expected $\pi/p$ for one of the SDC reconfigurable absorber
test-beam configurations \protect\cite{hangingfile}
in which $h/e\approx0.67$.  $m$, $E_{0\pi}$, and
$E_{0p}$ are from the fits shown in Fig.~11 of Paper~I.
}
\label{fig:pi_over_p}
\end{figure}

Via the ``universal spectrum'' concept we expect that 
the hadronic response of the calorimeter is the same for protons as for 
pions, except for a scale factor: $h/e$ is 
the same for both cases, and the mean hadronic fraction ratio $f^0_{\pi^-}/f_p^0<1$ is 
independent of energy.%
\footnote{It is implicit in this paper that $\pi^+$ and $\pi^-$ responses are 
essentially the same.
This was checked in a few cases, but was uninteresting. Since test beam work
invariably uses $\pi^-$ beams because they are free of $p$ contamination, charged
pions are simply labeled~$\pi^-$.} 
These statements are independent of a power-law approximation.  

In the power-law context, the energy-independent ratio $f^0_{\pi^-}/f_p^0$
should be $(E_{0\pi^-}/E_{0p})^{1-m}$. This  
means that $m$ is the same for both pions and protons.  The Paper~I 
(Fig.~11) simulations were consistent 
with equality.  The scale energy $E_{0\pi^-}$ was found to be about 1~GeV, with some
change from material to material.  For protons 
the Monte Carlo simulations yielded $E_{p0}\approx2.6$~GeV. 
These  were consistent with the expectation that $E_0$
was the approximate multiple-pion threshold\cite{tuscaloosa90}.  
Thus $f_{\pi^-}^0/f_p^0 \approx (1.0/2.6)^{1-m} = 
0.83$ for $m=0.815$ and $f_p^0/f_\pi^0 =0.87$ for $m=0.85$.

If $h/e\neq1$, a calorimeter should give a different response for 
charged pions than for protons. In the usual case, where $h/e<1$, pions 
give the larger response. The effect is illustrated in 
Fig.~\ref{fig:pi_over_p}, where as an example we use $h/e=0.67$, obtained 
from CALOR simulations (Paper~I, Table~1) for the ``iron" configuration of 
the SDC test-beam calorimeter\cite{hangingfile}. It is regrettable that 
there was not time to measure the effect there.

Equation~(\ref{eqn:pioverE_hadfrac}) may be rewritten for the pion and
proton cases:
\begin{eqnarray}
\pi/e&= 1-(1-h/e)f_{\pi^-}^0\nonumber\\
p/e &= 1-(1-h/e)f_p^0
\end{eqnarray}
Rearrangement gives us the energy-independent ratio
of the energy-dependent mean hadronic fractions:%
\footnote{This ratio was not calculated in Refs.~\cite{qcal97} or \cite{wigmans_pi_p_98}.}
\begin{eqnarray}
f_{\pi^-}^0/f_p^0 =& \frac{1-\pi/e}{1-p/e} \\
\label{eqn:pipratio}
\phantom{f_{\pi^-}^0/f_p^0}
\approx& (E_{0\pi^-}/E_{0p})^{1-m}
\label{eqn:pipratio_powerlaw}
\end{eqnarray}
The factor $(1-h/e)$ cancels.
The constant ratios of energy-dependent quantities
 given by Eq.~\ref{eqn:pipratio}  does not depend upon a power law or any other 
model for the hadronic fractions, although the statistical sensitivity is maximal for 
small $h/e$. 

Specialization to the power-law case results in Eq.~\ref{eqn:pipratio_powerlaw}.
Since only $a = (1/h/e)E_0^{1-m}$ can be found from $\pi/e$ (or $p/e$) measurements, the
scale energy cannot be found without assumptions about $h/e$. 
It is therefore interesting that the {\em ratio} of the scale energies given by 
Eq.~\ref{eqn:pipratio_powerlaw} can be very well known.

\begin{table}
\caption{Calculation of the hadronic fraction ratio
$f_{\pi^-}^0(E)/f_p^0(E)$ as a function of incident hadron 
energy (last column) using Eq.~(\ref{eqn:pipratio}) and data from Table~2 
of Akchurin et~al.\protect\cite{wigmans_pi_p_98}.  $\vev{f_{\pi^-}^0/f_p^0} 
=0.859\pm0.004$, with $\chi^2=13.2$. Most of $\chi^2$ is contributed by 
the first and third points.
\label{tab:wigmans}}
\vspace{0.4cm}
\begin{center}
\begin{tabular}{cccc}
\hline
\hline
Energy & \multicolumn{2}{c}{Response\cite{wigmans_pi_p_98}} & $f_{\pi^-}^0/f_p^0$ \\
\cline{2-3}
(GeV) & $p/e$ & $\pi/e$ \\
\hline
200 & $0.562\pm0.013$ & $0.647\pm0.001$ & $0.806\pm0.024$ \\
250 & $0.580\pm0.010$ & $0.648\pm0.001$ & $0.838\pm0.020$ \\
300 & $0.590\pm0.006$ & $0.658\pm0.001$ & $0.834\pm0.012$ \\
325 & $0.592\pm0.006$ & $0.652\pm0.001$ & $0.853\pm0.013$ \\
350 & $0.607^{+0.001}_{-0.004}$ & $0.659\pm0.001$ & $0.868\pm0.007$ \\
375 & $0.611^{+0.001}_{-0.003}$ & $0.664\pm0.001$ & $0.864\pm0.005$ \\
\hline
\hline
\end{tabular}
\end{center}
\end{table}

 \begin{figure}
\centerline{\includegraphics{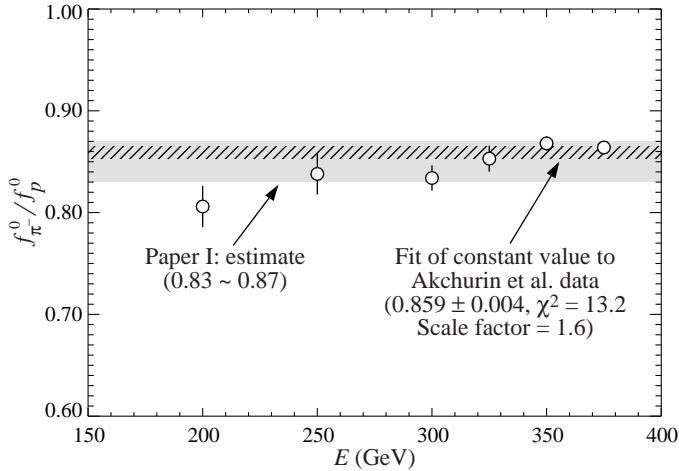}}
\caption{The mean hadronic fraction ratio 
$f_{\pi^-}^0/f_p^0$ as calculated from the copper/quartz-fiber 
calorimeter data of Ref.~\protect\cite{wigmans_pi_p_98}.  The gray band
is the range expected from Paper~I; the crosshatched band is a constant
value fitted to the data. The PDG scale factor is 1.6.
}
\label{fig:wigmans_pi_p}
\end{figure}

Since Paper~I was published, the CMS forward calorimeter group at CERN 
has measured the 
$\pi/p$ ratio using a calorimeter consisting of quartz fibers 
embedded in a copper matrix (QFCAL)\cite{qcal97,wigmans_pi_p_98}.
In this calorimeter, only  light was detected, most of it coming from em 
showers, so that $h/e$ was small, and the $\pi$--$p$ response 
difference was maximal. $\pi/e$ and $p/e$ as a function of energy are 
reported in Table~2 of Ref.~\cite{wigmans_pi_p_98} and copied to our 
Table~\ref{tab:wigmans}.

This ratio is calculated in the right column of 
Table~\ref{tab:wigmans} for the Akchurin et~al.\ data.  These data 
together with estimates of the Paper~I and a least squares fit to a 
constant $f_{\pi^-}^0/f_p^0$ are shown in Fig.~\ref{fig:wigmans_pi_p}.
The fit  yields $f_{\pi^-}^0/f_p^0\approx0.86$.

I am unable to find flaws in the several arguments leading to the conclusion
that $f_{\pi^-}^0(E)/f_p^0(E)$ is independent of energy, yet Fig.~\ref{fig:wigmans_pi_p}
shows evidence for energy dependence. A straight line with nonzero slope would
certainly fit the data better than a constant value. Akchurin et~al.\cite{wigmans_pi_p_98}
needed to make a careful but difficult subtraction of $\pi^+$ contamination in their 
positive beam. The contamination was minimal at the highest energy, as is 
reflected in the uncertainties. If the pion contamination correction were 
overdone, one would obtain the observed low values at the lower energies. 

In the case of incident kaons, the leading hadron is probably a strange meson, 
but sometimes a pion.  It is unlikely to be a proton or neutron.
The response difference between incident  pions and kaons should thus be small.

\section{\bold{mips}}
\label{sec:mips}

As indicated above and in Fig.~\ref{fig:EFlowSimple}, $e$ and $h$ are the
efficiencies with which electromagnetic and hadronic energy are converted
into a  visible signal. It is conventional to scale signal sizes, in ADC counts, 
to the mean response for minimum ionizing particles
 ({\em mips}), thus giving them something of an absolute meaning.  
 In practice the ``average'' signal from  penetrating muons, corrected for radiative losses,
 is assumed to be described by the Bethe-Bloch equation 
 including the density effect.%
 \footnote{In everyday detectors, energy loss by escaping $\delta$-rays or gain from 
entering $\delta$-rays is small (2\% level), so that  ``energy loss'' and ``energy deposit'' can be used
somewhat interchangeably.}
 It is then scaled to the value at
 minimum ionization, presumably defining the {\em mip}.
 
But the {\em mip} is commonly used incorrectly.

In a simplification of the normal derivation of the Bethe-Bloch formula,%
\footnote{Fano \cite{fano63} introduces an intermediate energy transfer region.}
ionization and excitation energy losses are calculated separately for (soft)
distant collisions (low energy transfer per interaction) and (hard) near 
collisions (high energy transfer). The regions are distinguished by the 
approximations appropriate to each\cite{fano63,rossi52,ADNDT01}.
One hopes for an energy at which they meet; this can sometimes be a problem 
in high-$Z$ materials.  Each contributes a factor $\ln\beta\gamma$ to the behavior at 
high energies:

\begin{enumerate}

\item 
As the particle becomes more relativistic, its 
electric field flattens and becomes more extended. This extension is  
limited by polarization of the material. This 
``density effect'' asymptotically removes the $\ln\beta\gamma$ factor 
contributed by the distant-collision region.  The ``relativistic rise'' is 
still there, but with half the slope\cite{RPP06}.%
\footnote{Review of Particle Physics 2006, hereafter RPP06.}

\item
The kinematic maximum energy $T_{\rm max}$ which
can be transferred in one collision sets the upper limit for hard energy transfer.  
Its rise with energy is responsible for  the other $\ln\beta\gamma$
factor.  As the particle energy increases there is more $\delta$-ray production and 
the ``Landau tail'' grows and extends.  The most probable energy loss, in a region well below 
minimum ionization that is dominated by many soft collisions, shows little or no
relativistic rise and approaches a ``Fermi plateau.''  This is more
easily understood for the related restricted mean energy loss discussed below.
\end{enumerate}
\jtem{2.2em}{}The $\delta$ rays are part of energy loss as described by the Bethe-Bloch
equation, and must not be confused with the density-effect correction $\delta(\beta\gamma)$ or
the muon radiative processes discussed below.

\begin{figure}
\centerline{\includegraphics[scale=.9]{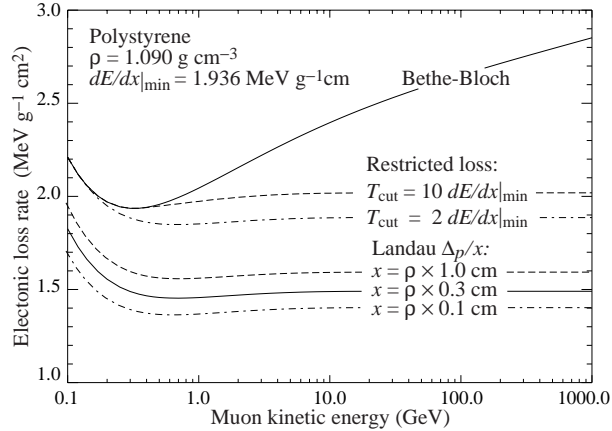}}
\caption{Bethe-Bloch $dE/dx$, two examples of restricted energy-loss rate, and the Landau most
probable energy deposit per unit thickness in polystyrene scintillator, in which 
$dE/dx|_{\rm min} = 1.936$~MeV~g$^{-1}\,$cm$^2$.  The incident particles are muons.
}
\label{fig:mulossscint}
\end{figure}

For detectors of moderate thickness $x$ (e.g., the scintillator tiles or 
LAr cells used in calorimeters),%
\footnote{$G \simle 0.05$--0.1, where $G$ is given by Rossi 
[\cite{rossi52}, Eq.~2.7.0].  It is Vavilov's $\kappa$\cite{vavilov57}.}
the energy-loss probability distribution 
$f(\Delta;\beta\gamma,x)$ is adequately 
described by the Landau (or Landau-Vavilov-Bichsel) 
disi\-tri\-bu\-tion\cite{vavilov57,bichsel88,landau44}.
The most probable energy loss is
\begin{equation}
\Delta_p = \xi\left[ \ln\frac{2mc^2 \beta^2 \gamma^2}{I} + \ln\frac{\xi}{I} + j -\beta^2 - 
\delta(\beta\gamma) \right] \ ,
\label{eqn:landau_most_prob}
\end{equation}
where $\xi = 0.153537\vev{Z/A}(x/\beta^2)$~MeV for a detector with a 
thickness $x$ in g~cm$^{-2}$, and $j=0.200$.%
\footnote{Rossi\cite{rossi52},
Talman\cite{talman79}, and others give somewhat different values
for $j$.  The most probable loss is not sensitive to its value.}
While $dE/dx$ is independent of thickness, $\Delta_p/x$ scales as 
$a\ln x + b$.
The density correction $\delta(\beta\gamma)$ 
was not included in Landau's or Vavilov's 
work, but it was later included by 
Bichsel\cite{bichsel88}.  It {\em must} be present for the 
reasons discussed in item (1) above. The high-energy behavior 
of $\delta(\beta\gamma)$ is such that
 \begin{equation}
\Delta_p \mathop{\longrightarrow}_{\beta\gamma{\simge}100} 
\xi\left[ \ln\frac{2mc^2\xi}{(\hbar\omega_p)^2} + j \right] \ ,
\label{eqn:landauasymptote}
\end{equation}
where $\hbar\omega_p$ is the plasma energy in the material, 21.8~eV in the case 
of poly\-styrene[RPP06, Tab.~27.1].
Thus the Landau most probable energy loss, like the restricted energy loss, 
reaches a Fermi plateau.   The Bethe-Bloch $dE/dx$%
\footnote{I follow convention and ignore the fact that $dE/dx$ is actually negative.}
and Landau-Vavilov-Bichsel $\Delta_p/x$
in polystyrene (scintillator) are shown as a function of muon energy in 
Fig.~\ref{fig:mulossscint}.  It is interesting that the asymptote is nearly reached at 10~GeV for muons, 
and that it is not much higher than the minimum at just under 1~GeV.  

In the case of restricted energy loss [RPP06 Eq.~(27.2)] the maximum kinetic energy transfer
in a single collision is limited to some $T_{\rm cut}\leq T_{\rm max}$. One may find the 
energy-weighted integral of the $\delta$-ray spectrum ($d^2N/dx dT$; RPP06 Eq.~(27.5) )
between $T_{\rm cut}$ and $T_{\rm max}$ to find that it is equal to the 
difference between the restricted and Bethe-Bloch
energy-loss rates.  Similarly, one can integrate the $\delta$-ray distribution over energy to find the
number  of $\delta$ rays produced in a tile---to find that in most cases $x\,dN/dx \ll1$.
As the incident particle energy increases, the tail
of the Landau distribution contains increasingly  energetic but improbable energy transfers.
Examples are shown in Fig.~\ref{fig:mulossscint}.

\begin{figure}
\centerline{\includegraphics{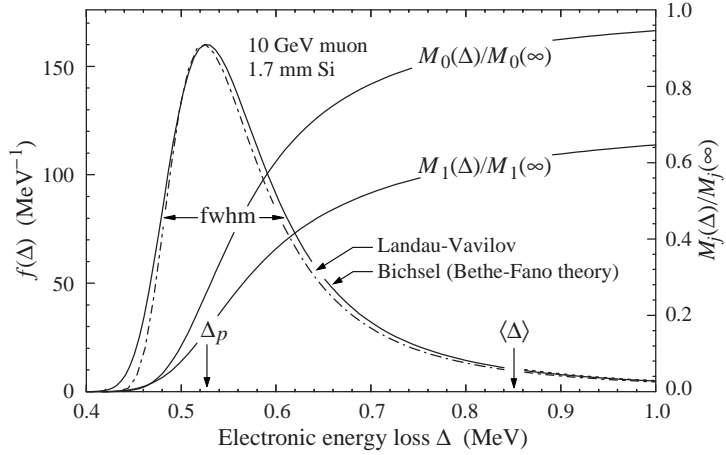}}
\caption{Bichsel's calculations of the electronic energy deposit distribution
for a 10 GeV muon traversing a 1.7~mm silicon detector
(which has roughly the stopping power of a 3-mm thick 
scintillator)\cite{bichsel88,bichsel06pc,bichselAMOPH,bichsel06}.
The Landau-Vavilov function (dot-dashed) uses a Rutherford 
cross section without atomic binding corrections but with a maximum
kinetic energy transfer limit $T_{\rm max}$. 
The solid curve was calculated using Bethe-Fano theory.  
$M_0(\Delta)$ and $M_1(\Delta)$ are the cumulative 0th and 1st 
moments of $f(\Delta)$, respectively. The fwhm of the Landau-Vavilov
function is about $4\xi$ for detectors of moderate thickness.  $\Delta_p$ is the most probable
energy loss, and $\vev{\Delta}$ divided by the thickness is the Bethe-Bloch mean, $dE/dx$.
}
\label{fig:bichselthickSi}
\end{figure}

In summary: The mean of the energy loss given by the Bethe-Bloch equation
is ill-defined experimentally and is not useful for describing
energy loss by single particles. \footnote{``The expression $dE/dx$ should be
abandoned; it is never relevant to the signals in a particle-by-particle analysis.''\cite{bichsel06}}  
(It probably finds its best application in dosimetry,
where only bulk deposit is of relevance.)  
It rises with energy because $T_{\rm max}$
increases. The large single-collision energy transfers
that increasingly extend the long tail are rare.  

For a $\beta\approx1$ particle, 
for example, on average only one collision with $T>1$~keV 
will occur along a path
length of 90~cm of Ar gas\cite{bichsel06}. 
The energy-loss distribution for a 10~GeV muon traversing 
a 1.7~mm silicon detector, shown 
in Fig.~\ref{fig:bichselthickSi}, further illustrates the point.
Here about 90\% of the area ($M_0(\Delta)/M_0(\infty)$) 
but only $\sim 60\%$ of the energy deposition 
($M_1(\Delta)/M_1(\infty)$) falls below the Bethe-Bloch $\vev\Delta$,
and at this energy  $\Delta_p \approx 0.6\vev\Delta$.  The long
tail of $f(\Delta)$ extends to $T_{\rm max} = 4800$~MeV.

The mean of an experimental sample consisting of a few hundred events is subject
to these large fluctuations, sensitive to cuts, and sensitive  to background.  
The mean $\vev{\Delta}_{\rm exp}$ deduced from
the data will almost certainly underestimate the true mean; 
in general $\Delta_p < \vev{\Delta}_{\rm exp} < \vev{\Delta}$. 
On the other hand, fits to the region around the stable peak in the pulse-height distribution 
provide a robust determination of the most probable energy loss.  The peak is somewhat
increased from the  Landau $\Delta_p$ by experimental resolution function.
 
The ionization/excitation energy fraction sampled by the active region is 
conventionally given by
\begin{equation}
\hbox{fraction sampled} = \frac{S\,dE/dx|_{\rm scint}} {S\,dE/dx|_{\rm scint} +
A\,dE/dx|_{\rm abs}} \ .
\end{equation}
Boundary-crossing $\delta$ rays are (accurately) assumed not to be important.
Here $S$ is the active region thickness, $dE/dx|_{\rm scint}$ is the energy
loss rate in the active region (scintillator or other), $A$ is the absorber thickness, 
and $dE/dx|_{\rm abs}$ is the energy-loss rate in the absorber.  It is ``fairly sampled'' if 
this is the case.  It is not the case for electron and muon radiative losses.

Relativistic muons also lose energy radiatively by direct pair production,
brems\-strahlung,  and photonuclear interactions.  In iron at 1~TeV, these loss rates are
in the ratios 0.58:0.39:0.03. The ratios are fairly insensitive to energy at energies
where the radiation contribution is important. The pair:brems\-strah\-lung ratio is about the same
from material to material, while the photonuclear fraction grows with atomic number.
These contributions to $dE/dx$ rise almost linearly with energy, becoming 
as important as ionization losses at some ``muon critical energy'' $E_{\mu 
c}$: 1183~GeV in plastic scintillator, 347~GeV in iron and 141~GeV in lead.%
\footnote{Other charged particles 
experience radiative losses as well, but there is no easy mass scaling for
the radiative loss rate.  In a calorimeter incident high-energy pions lose energy by both radiation 
and ionization until they interact, but the higher loss rate is of little consequence and 
in any case the radiated energy is absorbed.}

The tables of Lohmann et~al.\cite{lohmann85} are commonly 
used.  More extensive tables with a somewhat improved treatment of 
radiative losses are given by Groom et~al.\cite{ADNDT01}, and an extension 
to nearly 300 materials is available on the Particle Data Group web 
pages\cite{AtomicNuclearProperties}.%
\footnote{For the PDG tables, an 
improvement to the pair-production cross section was made 
which slightly changes the muon $dE/dx$ at high energies in high-$Z$ 
materials.  In both cases the ionization losses include a correction for
muon bremsstrahlung on atomic electrons, so that at the highest energies the table
entries slightly exceed the Bethe-Block values.}

Muons in an absorber are  accompanied by
an entourage of photons and electrons (cascade products from direct pair production and 
bremsstrahlung) characteristic of radiative losses in the higher-$Z$ absorber.  
If the calibration muon beam has been momentum-selected and then travels 
through air or vacuum to the calorimeter, it enters ``naked,'' without its entourage 
of pairs and bremsstrahlung photons and, until this builds up over several 
radiation lengths, the signal distribution does not include the full radiation 
contribution.  If the equilibrium contribution is desired,  absorber should be placed 
in the beam.  In principle these radiative products should 
be detected with the same efficiency as electrons, although a few high-energy
pairs  go through the active layers.  There is little radiative loss in the low-$Z$ 
active layers themselves.

Bremsstrahlung is sufficiently continuous as to not introduce significant
radiation fluctuations\cite{vanginneken86}, but there are large fluctuations
in the pair production energy loss.  Monte Carlo calculations by Striganov and 
collaborators\cite{striganov06}
indicate that, while a high-energy shoulder appears on the energy-loss distribution, 
the most probable energy loss increases only slightly in ``thin'' absorbers,''  e.~g.~for
1000~GeV muons incident on 100~g~cm$^{-2}$ of iron.  They regard radiative effects
as ``important'' when the most probable height of the normalized energy-loss
distributions are lowered by $\simge10$\% when radiative effects are included.  
This is the case for the total signal from real calorimeters (more than
1000~g~cm$^{-2}$) at the highest muon calibration energies normally used. 
Although the most probable energy loss is still the best calibration metric, 
it does rise somewhat with beam energy because of the radiative effects.

The calibration of the HELIOS modules (uranium/scintillator sandwiches)  is particularly well 
described\cite{akesson87}.  The common normalization of individual layers was done 
via radioactive decay in the uranium plates, so the distributions shown are for entire modules.
In correcting for the radiative losses they assumed 
fair sampling by the scintillator.  Their correction of 
energy deposit for radiative effects is straightforward, but the robustness of 
the ``average'' energy deposit is unclear.

Muon detection in SPACAL (scintillator fibers in a lead matrix) is carefully described by 
Acosta et~al.\cite{spacal92}.   The energy dependence of the most 
probable values is shown.  The distributions clearly show both the radiative
broadening and the increase of the most probable values due to 
radiative effects.

In both of these cases, the calorimeters as a whole were calibrated. 
Since these are  many radiation lengths thick, the lower average deposit 
at the beginning should not sensibly affect  the result.

All of this assumes muons of known energy.  ``Out of channel'' muons, 
which have gone through or around the test-beam optics, are certainly not 
dependable calibration particles, but are sometimes used\cite{hangingfile}.  
Cosmic ray muons 
have a characteristic energy of about 3~GeV, but the flux falls off 
as about $\cos^2\theta$, where $\theta$ is the zenith angle [RPP06, Sec.~24].  
They can provide a useful if imprecise calibration in some situations.

\section{\bold{e}, \bold{h}, and \bold{e}/\bold{h}}\label{sec:e_h}

Given a credible muon calibration, the quantity $e/mip$ can be measured in an 
electron beam.  In a sampling calorimeter, cascade electrons are 
predominately produced and absorbed in the inactive higher-$Z$ material, 
so the signal is significantly smaller than might be expected from the 
active layer's share of $dE/dx$ ($e/mip\approx 0.6$ to 
0.7\cite{wigmans87}), but with uncertainty associated with {\em most probable} 
energy deposit vs {\em average} energy deposit. It can be ``tuned'' by changing the 
absorber/detector ratio, perhaps to achieve compensation ($h/e =1$). Other 
things being equal, the detection efficiency is smaller if the absorber 
has a higher $Z$. The critical energy is lower, so characteristic shower 
electrons are more likely to deposit their energy before leaving the 
absorber.  As a corollary, $e/mip=1$ for a nonsegmented calorimeter (e.g., an 
inorganic crystal), and since there is always missing hadronic energy 
such a calorimeter is always noncompensating.

The hadron efficiency $h$ is more problematical; one finds $a = 
(1-h/e)E_0^{1-m}$ or an equivalent by fitting the energy dependence of 
$\pi/e$, and assumptions must always be made about the constant 
multiplier to find $h/e$ and hence $h/mip$.  The multiplier $E_0^{1-m}$ is close to 1 
for incident pions, but it is about 20\% higher for protons.
$h$ is considerably more difficult to model, but in general it is smaller 
than $e$ because of the wide variety of ways hadronic energy becomes 
invisible, e.g., through nuclear binding energy losses and ``late" energy 
deposition (outside the electronics window).\footnote{A particularly nice 
discussion is given by Ferrari and Sala\cite{ferrari00}. } It increases 
somewhat with $Z$, and can be enhanced by neutron production in uranium. 
It can also be ``tuned" by the choice of material and sampling fraction\cite{wigmansbook}.

Can we measure $h/mip$ directly?  Only by observing hadronic cascades in a 
calorimeter insensitive to $\pi^0$-produced em cascades, or by observing 
cascades produced by hadrons below the $\pi^0$ threshold.  In Paper~I we 
speculated about building a calorimeter sensitive only to hadrons (a 
neutron detector) or to the em sector (a calorimeter sensitive only to 
Cherenkov radiation), but the context of the discussion was verification
of the power-law approximation for $f^0_h$ and determination of the power~$m$.

In the spirit of only- (mostly-) em sensitivity in the copper/quartz-fiber CMS test
calorimeter, Demianov et~al.\cite{bonnersphere} made preliminary neutron 
measurements using Bonner 
spheres\cite{bonner60} adjacent to the copper/quartz-fiber 
calorimeter\cite{qcal97}.  The longitudinal and transverse 
distributions were measured.  The results were in fair to good agreement 
with MARS96\cite{MARS96} calculations, but not sufficiently detailed to 
obtain $h/mip$ (or $n/mip$).  Preliminary proposals\cite{ILCdual} 
(in connection with International Linear Collider (ILC) detector R\&D) are being made to 
measure the neutron flux by a variety of methods; future test-beam results will be of 
great interest.  The problem will be discussed at more length in Sec.~\ref{sec:devil}.

One might use hadrons with energies below the $\pi^0$ threshold. ZEUS
collaborators made measurements with low-energy protons and charged pions 
with a compensated U/scintillator calorimeter\cite{zeus_proton}. 
Interestingly, as the kinetic energy of the beam was decreased from 
about 5~GeV to about 0.4~GeV, $e/h$ decreased from its high-energy value (one) to the 
$e/mip$ measured for electrons.  The lower-energy particles tended to lose much or all of 
their energy by ionization, so they became indistinguishable from 
electrons at sufficiently low energies. The resolution also decreased from 
its hadronic value, approaching the em resolution until at the lowest 
energies noise became dominant.

A more desirable (or complimentary) approach might  be to use an incident 
beam of low-energy neutrons. Since $E_{0p} \approx 2.6$~GeV, one might 
expect the $\pi^0$ threshold to be about $T \approx 1.6$~GeV. As the 
energy is scanned downward, a pure hadronic signal should emerge.  The 
response would not be quite the hadronic signal observed from a 
higher-energy cascade, but this difference can probably be understood.
At very least, measurements in a low-energy neutron beam would be interesting.
The real problem is making the test beam.

\section{Resolution\label{sec:resolution}}

The arrows between boxes in Fig.~\ref{fig:EFlowSimple} actually
indicate the various p.d.f.'s describing fluctuations in
each of the steps. A more complete version, Fig.~\ref{fig:EFlow_all}, defines
these distributions, which are described in more detail in 
Table~\ref{tab:resfunctions}.
In the simple model considered in this paper, five p.d.f.'s appear. 

The {\em potentially}  detectable energy deposit, or visible energy,
is labeled ``vis." It usually means ionization in a sensing medium, such as scintillator
or liquid argon.  
In the rare cases where  Cherenkov light is to be sampled, it means the Cherenkov radiation
produced.  It contains the variations due to energy deposit, not detection.
The variance associated with the visible energy distribution
at fixed $ E_{\pi^0}$, dominated by fluctuations in the
total kinetic energy of neutrons, is the {\em intrinsic} variance.

This ionization is then {\em sampled}  
directly but more often via scintillators, where the scintillation
light is usually detected by photomultipliers.  
The label ``samp'' refers to the additional fluctuations introduced in 
this process.

The stochastic processes are defined as follows:

\begin{enumerate}

\item
In the cascade initiated by the incident hadron, energy $E_{\pi^0}$ is
transferred to the $\pi^0$ sector via $\pi^0$ production and decay.  Because of
its different energy dependence, em energy deposited by nuclear gamma rays
is not included in my definition of $E_\piz $.  The p.d.f.\ $\Pi(E_\piz )$ was
introduced in Sec.~\ref{sec:albedo} to describe its distribution. 

\item
The $\piz$\ energy $E_\piz $ is detectable with some average efficiency $e$ via the
ionization produced mostly by low-energy electrons and
positrons.  The p.d.f.\ of the em visible energy deposit at fixed $E_\piz $ is
labeled $g_\piz(E^{\rm vis}_\piz|E_\piz )$.

\item
Quite independently, the hadronic component produces a visible signal via
energy loss by charged secondaries.  This signal, the result of a variety
of mechanisms, is produced with an overall average efficiency $h$.  The  
visible contribution by nuclear gamma rays is included here.  Again, most
of the ionization is by low-energy particles. The p.d.f.~$g_h(E^{\rm      
vis}_h|E_\piz )$ describes the distribution at fixed $E_\piz$.

\item
Only the total deposit $E^{\rm vis} = E^{\rm vis}_\piz + E^{\rm vis}_h$ can be detected.
Convolution over the intermediate variable $E^{\rm vis}_\piz$ yields the
p.d.f.\ $F_{\rm vis}(E^{\rm vis}|E_\piz )$.   The product
of this distribution with $\Pi(E_\piz )$, integrated over $E_\piz $, is the visible energy 
distribution which can be sampled.  The width of this distribution is identified with 
the ``intrinsic resolution.''

\begin{figure}[t]
\centerline{\includegraphics{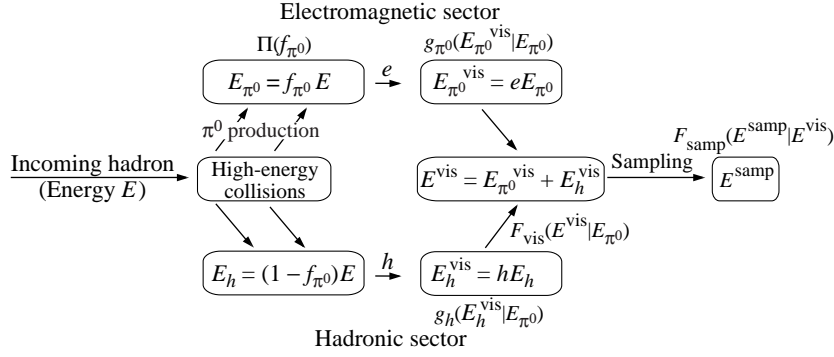}}
\caption{Energy flow in a calorimeter, with the statistical distributions
contributing to the experimental resolution indicated.
}
\label{fig:EFlow_all}
\end{figure}

\item
Finally, the visible energy is sampled by measuring the ionization,
either directly or by observing scintillation light with photomultipliers or photodiodes.%
\footnote{The observation of Cherenkov light is another option.} 
This step is to an extent under the 
control of the experimenter, since it depends on scintillation efficiency,
light collection efficiency, and other design details.  For fixed $E^{\rm
vis}$ one measures a signal $E^{\rm samp}$, chosen from a distribution
$F_{\rm samp}(E^{\rm samp} | E^{\rm vis})$, which is then summed over the 
intermediate $E^{\rm vis}$ to obtain the final distribution of the signal,
$E^{\rm samp}$. Even this step is not a simple convolution, since the
variance of $F_{\rm samp}(E^{\rm samp} | E^{\rm vis})$ is proportional to 
$E^{\rm vis}$, not~$E$.

\end{enumerate}

The intrinsic and sampling distributions were separated in a classic
experiment by Drews et~al.\cite{drews90}, who used compensating
sandwich calorimeters with scintillator readout with either lead or uranium
plates. Alternate sets of scintillators were read out
separately.  Sampling variations in the two sets were independent, while
intrinsic fluctuations were correlated.  These were recovered by adding
and subtracting variances.

Calculation of the combined distribution is tedious and not entirely
obvious; the details are relegated to Appendix~A. The result (repeating
Eq.~(\ref{eqn:resolution})) is
\begin{equation}
\left({\sigma\over E}\right)^2 =
\frac{(\pi/e) \sigma_{\rm samp0}^2 }{E}
+ \left[ \frac{f_{\pi^0}^0 \sigma_{e0}^2 }{E}+ \frac{f^0_h \sigma_{h0}^2 h/e}{E}\right]
+ (1-h/e)^2\sigma_\piz^2(E) \ .
\label{eqn:resolution_text}
\end{equation}
Here $\sigma^2_{\rm samp0}$, $\sigma^2_{e0}$, and $\sigma^2_{h0}$ 
scale the variances contributed by the sampling, $\pi^0$ energy deposit, 
and hadronic energy deposit, respectively.   They have the units of energy.

The first term is the familiar sampling contribution,
except that it is multiplied by $\pi/e$.  This is to be expected {\em and
required}, since this contribution to the variance is proportional to the
sampled visible signal, with mean $(\pi/e)E$, rather than to the incident
energy $E$.

The two terms in the square brackets are the two pieces of the intrinsic
variance. Even if $h/e=1$, the intrinsic variance has some energy
dependence, since $f_{\pi^0}$ increases with energy and $f_h$ decreases  
with energy.

Together, the sampling term and the two intrinsic terms in the square brackets are usually 
represented as $(C/\sqrt{E})^2$, ignoring the energy dependence of each of the three terms.

\begin{figure}[t]
\centerline{\includegraphics{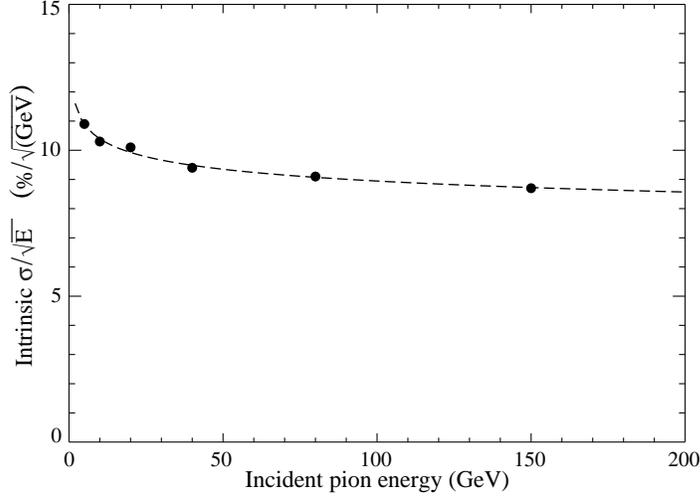}}
\caption{Wigman's (simulated) data\cite{wigmansbook} for the intrinsic 
resolution of a Pb/LAr calorimeter scaled by $1/\sqrt{E}$, fitted with
the term in the square brackets of Eq.~(\ref{eqn:resolution_text}).  
The fit parameter are given in the text.
}
\label{fig:wigmans_intrinsic}
\end{figure}

Wigmans\cite{wigmansbook} has noted that $\sigma_{\rm intr}/\sqrt{E}$ for
the simulated lead/LAr calorimeter described in his Table~3.4 decreases
with energy, reflecting the gradually increasing transfer of energy to the
$\piz$ sector.  His calculated results for six energies, given in his
Table~4.3, are plotted in Fig.~\ref{fig:wigmans_intrinsic}.  My curve was
obtained by adjusting the intrinsic variance scales $\sigma^2_{e0}$ and
$\sigma^2_{h0}$. The best-fit parameters are $\sigma_{e0} = 5.1\%$ and   
$\sigma_{h0} = 13.7\%$.  The fit is remarkably good, and, as expected,
$\sigma_{h0}$ is considerably larger than $\sigma_{e0} $.

The last term is the expected ``constant term.''  Its mild energy dependence is discussed in
Sec.~\ref{sec:albedo}.  It must approach zero at high energies ($1/E\to0$), as $\Pi(f_\piz)$ ``crowds against"
the $f_\piz=1$ limit.  That it can be represented as a constant is an artifact
of the limited energy range of test-beam measurements. This point is discussed in more detail
below.

\begin{figure}[t]
\centerline{\includegraphics{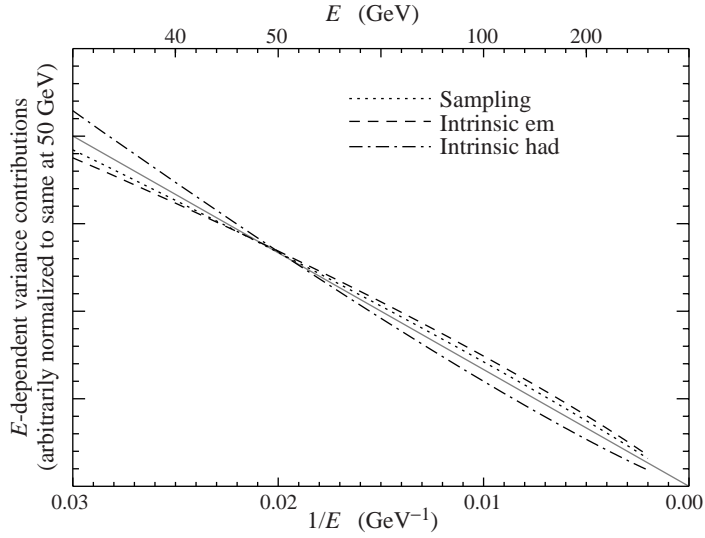}}
\caption{Shapes of contributions to sampling and intrinsic variance.  As
$1/E\to0$, the slopes of the sampling and $\piz$contributions
approach finite constants, since $\pi/e\to1$ and $f_{\pi^0}\to1$, while 
the slope of the intrinsic hadronic contribution approaches zero
($f_h\to0$).
}
\label{fig:groom_res_shape}
\end{figure}

\begin{figure}[t]
\centerline{\includegraphics{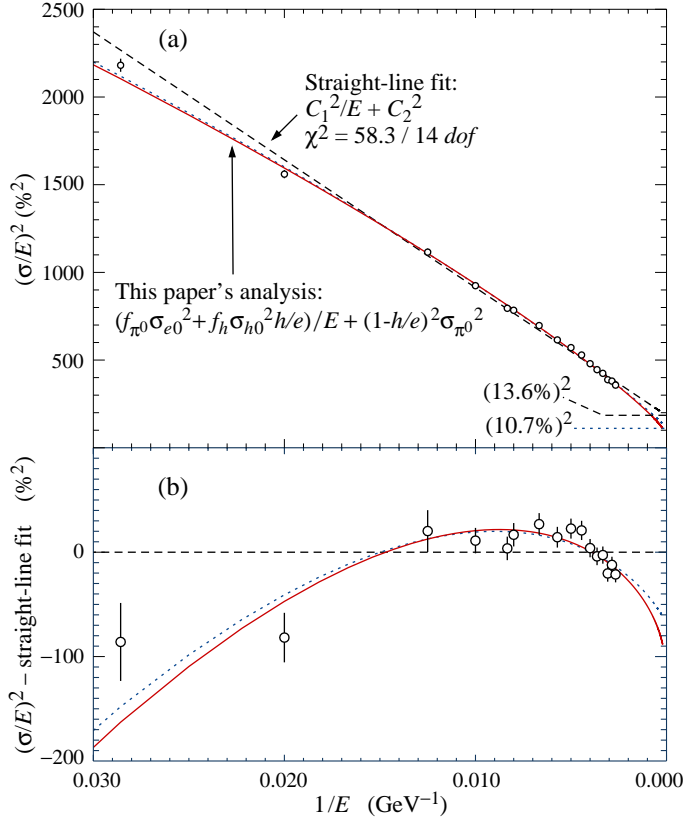}}
\caption{``Conventional'' and Eq.~(\ref{eqn:resolution_text}) fits to the
copper/quartz-fiber test module energy response data given in Table 3 of
Ref.~\cite{qcal97}.  The fit parameters are given in Table~\ref{tab:fitparameters}.
The fit shown by the dotted curve is for constant  $\sigma_\piz$, while the solid curve fit
is for a more physically reasonable weak negative power law $E$-dependence.  The high-energy
``turndown'' is more clearly shown in (b), where the fits and data are shown relative to 
the usual straight-line fit.}
\label{fig:groom_final_resplot}
\end{figure}

The $\pi^0$ part of the intrinsic variance fraction increases in importance as $E$
increases, as does the sampling variance. Both curve downward if plotted
vs $1/E$, as shown in Fig.~\ref{fig:groom_res_shape}. The hadronic
intrinsic contribution curves upward, since it decreases faster than $1/E$.

It is difficult to verify Eq.~(\ref{eqn:resolution_text}), even with
robust experimental data. The expected resolution should be a linear
combination of the three curves shown in Fig.~\ref{fig:groom_res_shape}
(plus a constant term), so any deviation of the variance from the   
traditional $C_1^2/E$ will show up as a slight curvature.  Moreover, the
sampling and $\pi^0$ contributions have such similar energy dependence that a
simultaneous fit to $\sigma_{\rm samp0}$ and $\sigma_{e0}$ can be
indeterminate.

\begin{table}
\caption{Parameters for the three fits shown in Fig.~\ref{fig:groom_final_resplot}.
The power-law parameters $m=0.833$, $a=0.753$  and $h/e=0.247$ (for
$E_0=1.0$) from Table~\ref{tab:pioe} were used in the reduction.  The experimental 
data are from the 3rd ($\sigma_{\rm rms}/E)$ column of Tbl.~3 in Ref.~\cite{qcal97}.  }
\vspace{0.4cm}
\begin{center}
\begin{tabular}{cc c cc c cc}
\hline
\hline
\multicolumn{2}{c}{$\sigma/E=C_1/\sqrt{E} \oplus C_2$} & 
   &\multicolumn{2}{c}{Eq.~(\ref{eqn:resolution_text}) with 
   $\sigma_{\rm samp0} = 0$} &&\multicolumn{2}{c}{Eq.~(\ref{eqn:resolution_text}) with 
   $\sigma_{\rm samp0} = 0$} \\
\cline{1-2}\cline{4-5}\cline{7-8}
Parameter &Value &&Parameter &Value&&Parameter &Value\\
\hline
&&&$\sigma_{e0}$ & 377\% &&$\sigma_{e0}$ & 372\%\\
$C_1$ & 270\% &&$\sigma_{h0}$ & 216\%&&$\sigma_{h0}$ & 214\%\\
&&&$s_1*$&14.2\%&&$s_1*$ & 15.7\%\\
&&& $s_2*$ & 0 (fixed)&& $s_2*$ & $-$0.058\\
$C_2$ & 13.6\%&&$(1-h/e)s_1 $ & 10.7\%\\
$\chi^2/dof$ & 58.3/14 && $\chi^2/dof$ & 18.6/13&& $\chi^2/dof$ & 18.0/12\\
\hline
\hline
\multicolumn{8}{l}{* $\sigma_\piz = s_1\, (E/100\,{\rm GeV})^{s_2}$} \\
\label{tab:fitparameters}
\end{tabular}
\end{center}
\end{table}

The square of the fractional energy resolution in the copper/quartz-fiber
test calorimeter for incident pions (Akchurin et~al.\cite{qcal97}, Table
3) is plotted as a function of $1/E$ in Fig.~\ref{fig:groom_final_resplot}. The
data curve downward relative to the ``conventional'' linear fit, $C_1^2/E
+ C_2^2$, shown by the dashed line. For this calorimeter intrinsic  
fluctuations were more important than sampling fluctuations except at the
lowest energies, although sampling fluctuations were not negligible.
Because of the nearly-degenerate shapes  of the sampling and intrinsic $\pi^0$ 
fluctuation curves,
I set $\sigma_{\rm samp0}=0$ in making a fit, which is shown by the solid
curve in the Figure.  It describes the data well, and the physics
responsible for the curvature is understood.

Parameters for both cases are shown in Table~\ref{tab:fitparameters}.     
The fitted values for $\sigma_{e0}$ and $\sigma_{h0}$ are very much
larger than for the example discussed above and shown in
Fig.~\ref{fig:wigmans_intrinsic}; this follows from the excellent
resolution of Wigman's model calorimeter and the (by design) poor   
resolution of the copper/quartz-fiber calorimeter.

Other examples testing the $1/E$ dependence are hard to find.  Many test-beam
results are at low energy, many have large errors, and many of the earlier
results are presented as functions of $C_1/\sqrt{E} + C_2$ rather than  
$C_1/\sqrt{E} \oplus C_2$.  It will be of interest to test
Eq.~(\ref{eqn:resolution_text}) against further experimental results.

\section{Jets}
\label{sec:jetresponse}

How is calorimeter's response to a jet different than its response to a 
single pion? There are three situations to consider:
 
\begin{enumerate}

\item
An incident pion. The primary collision usually occurs about an 
interaction length into the calorimeter. There is a minimum of backscatter 
(``albedo'').  The fragmentation process is dependent on energy and the 
nuclear environment.

\item
A ``test-beam jet,'' in which trigger counters ensure that the primary 
interaction occurs in a thin absorber in front of the calorimeter. This is 
exactly the same as the incident pion case, except for increased albedo 
because of the high probability that some of the first-collision debris 
interacts near the front of the calorimeter.

\item
A primary fragmentation jet.  The only evident differences from the above 
cases are the (much) higher energies and a simpler environment; except in 
heavy-ion collisions just two particles interact. This section concerns 
whether the mix and distribution of photons, pions, and other particles 
results in calorimeter response different than the response to a pion or 
``test-beam jet.''

\end{enumerate}

As elsewhere in this paper, the situation is highly idealized: The 
homogeneous or fine-sampling calorimeter is large enough to contain the 
entire cascade and the structure is uniform throughout. The realities of 
jet-finding and isolation algorithms, albedo, the effects of the magnetic 
field, passive material in front of the calorimeter, etc., are all 
ignored.

The power law approximation for $f_h$ developed in Paper~I will be used 
throughout this section.

A jet with energy $E_J$ consists of photons, mostly from $\pi^0$ decay, 
and ``stable" hadrons. (Energy which might be carried away by leptons is 
ignored.) Since most of the incident ``stable'' hadron flux consists of 
charged pions, $E_0\approx1$~GeV and $(1-h/e)\approx a$.

One needs only to sum the calorimeter response to all of these particles 
to obtain the response to a jet. If $R_{\piz j}$ is the response to the $j$th 
$\pi^0$ (with energy $E_{{\pi^0}j}$) in the jet and $R_{hk}$ the 
response to the $k$th stable hadron (with energy $E_{hk}$), then the 
response to a jet is given by
\begin{equation}
E_J^{\rm vis} = \sum_{j=1}^{N_\piz} R_{\piz j}+ \sum_{k=1}^{N_{\rm had}}
    R_{hk} \ .
    \label{eqn:Ejetsummation}
\end{equation}
Using Eqns.~(\ref{eqn:e_response})--(\ref{eqn:pioverE}) to evaluate 
$R_{hk}$ and $R_{\piz j}$, this reduces to
\begin{equation}
E_J^{\rm vis}=e E_J\Big[1 -a\, E_J^{m-1}
\sum_{k=1}^{N_{\rm had}} (E_{hk}/E_J)^{m} \Big]  \ .
\label{eqn:jetsummation}
\end{equation}

Alternatively, the spectrum of stable hadrons can be described by a 
fragmentation function $D(z)$, where $z$ is the hadron's momentum parallel 
to the jet direction, scaled by the jet's momentum. In the present study 
$z$ is treated as the fractional energy, i.e.\ $z\approx E_{\rm had}/E_J$. 
When the arguments leading to Eq.~\ref{eqn:jetsummation} are repeated, 
one obtains
\begin{equation}
E_J^{\rm vis} =e E_J\Big[1 -a\,E_J^{m-1}
  \int_0^1 z^mD(z)dz \Big] \ ,
  \label{eqn:jetintegral}
\end{equation}
where $D(z)$ describes the spectrum of all hadrons except for the 
$\pi^0$'s.

The sum (Eq.~\ref{eqn:jetsummation}) or integral 
(Eq.~\ref{eqn:jetintegral}) thus appears as a correction factor to the 
normal hadronic response of a calorimeter.  {\em If it is unity, the 
response to a jet is the same as the response to a pion.} In any case it 
is multiplied by $a$, which is usually $<0.3$. The distinction between a 
single pion and a jet vanishes as the calorimeter becomes more 
compensated---except, of course, for the albedo, magnetic field, passive 
material in front of the calorimeter, and cone-cut effects mentioned 
above.

If the sum or integral is evaluated for $m=0$, the mean stable hadron 
multiplicity $\vev{N_{\rm had}}$ is obtained. If $m=1$, the result is the 
mean nonelectromagnetic fraction of the jet's energy $\vev{F_{\rm had}}$. 
The desired summation or integral, with $m\approx0.82$--0.86, is in some 
sense an interpolation between the two.

In using either experimental or Monte Carlo distributions to evaluate the 
sum or integral, special treatment of the very low-$z$ region is 
necessary, as is normalization to an appropriate $\vev{F_{\rm had}}$.

The integral in Eq.~\ref{eqn:jetintegral} is evaluated for four 
representative cases:

Two experimental results, both with jet energies at or near $M_Z/2$. Since 
the measurements are for {\it charged\/} hadrons, the distributions must 
be renormalized to include the contributions of such particles as 
$\Lambda$'s and $K_L$'s.

\begin{enumerate}

\item
Jets from $Z$ decay, as measured by the DELPHI collaboration at 
LEP\cite{delphi}.  The published fragmentation function is for the entire 
event, so the function has been normalized downward by a factor of two to 
describe the individual jets. Data were read from their Fig.~3(b) and 
extrapolated to $z=0$.

\item
CDF charged fragmentation function at $\sqrt{s} = 1800$~GeV\cite{CDFjets}. 
$z dN_{\rm ch}/dz$ was extrapolated to $z=0$ to force $\vev{F_{\rm ch}} = 
0.65$, their reported value. (Since some of the energy is carried by 
neutrals, this value is probably too high for consistency with isospin 
conservation.)

\end{enumerate}

\begin{figure}
\centerline{ \includegraphics{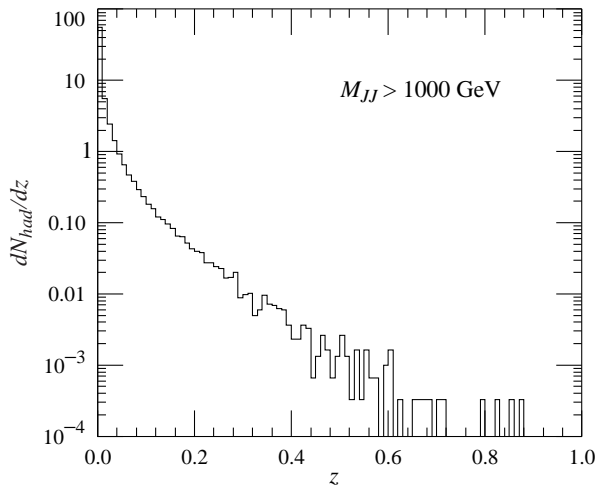}} 
\caption{Distribution in $z$ for ISAJET TWOJET events at $\sqrt{s} =
40$~TeV, for $1000\ \hbox{GeV}\ < M_{JJ} <2000\ \hbox{GeV}$.
}
\label{fig:dnhdz1000}
\end{figure}

Two samples of TWOJET ISAJET\cite{ISAJET} events at $\sqrt{s} = 40$~TeV.%
\footnote{I am indebted to my SDC collaborator E.~M. Wang for running 
these simulations.  This work was jointly reported in 
Refs.~\cite{fortworth90} and \cite{aachen90}.}
In both cases, all hadrons other than $\pi^0$'s are used:

\begin{enumerate}
\setcounter{enumi}{2}

\item
3226 events with $p_t$ (hard scatter) $>40$~GeV/$c$, and $(100\ \hbox{GeV} 
< M_{JJ} <200\ \hbox{GeV})$.  The mean jet momentum is 73~GeV/$c$, and the 
mean non-$\pi^0$ hadronic multiplicity is 26.

\item 
3042 events with $p_t$ (hard scatter) $>400~{\rm GeV/}c$, and $1000\ 
\hbox{GeV} < M_{JJ} <2000\ \hbox{GeV}$.  The mean jet momentum is 
677~GeV/$c$, and the mean non-$\pi^0$ hadronic multiplicity is 70.  The 
$z$ distribution for these events is shown in Fig.~\ref{fig:dnhdz1000}.

\end{enumerate}

The results are summarized in Table~\ref{tab:fragmentation}. There is 
ambiguity because of uncertainty in $\vev{ F_{\rm had} }$ in the 
simulations and $\vev{ F_{\rm ch} }$ in the experimental results.  If pion 
production dominates, one might expect $\vev{ F_{\rm had} }\approx2/3$ 
from isospin considerations. (In Paper~I, we reported fractions closer to 
3/4.) Some of the bias can probably be removed by normalizing $\int z D(z) 
dz$ to 2/3, as indicated by the table entries in parentheses.  As can be 
seen, the integral is slightly less than unity for the similar low-energy 
LEP and Tevatron fragmentation functions, and it is slightly greater than 
unity for simulated 40~TeV jets.  Values lie between 0.84 and 1.15 before 
normalization, and 0.92 to 1.06 after normalization---probably well within 
the uncertainty of the fragmentation functions in either the experimental 
or Monte Carlo cases.  The integrals also change by about 0.05 if $m$ is 
changed by 0.01, introducing an additional uncertainty which could be as 
great as 20\%. Given the various uncertainties, I conclude that the 
correction factor for fragmentation jets at the highest-energy colliders 
should be between 0.85 and 1.15.

\begin{table}
\caption{Integrals over representative fragmentation
functions. Numbers in parentheses are calculated for the
nonelectromagnetic energy fraction normalized to 0.67. In the case of
the DELPHI and CDF results, the unrenormalized energy fraction is for
charged hadrons only.\label{tab:fragmentation}}
\begin{center}
\begin{tabular}{l l ccc}
\hline
\hline
Source&Process &$\int_0^1D(z)dz$&$\int_0^1z^{0.86}D(z)dz$
   &$\int_0^1zD(z)dz$ \\
\hline
DELPHI&$Z\to jet\ jet$&11.0(12.1)&0.84(0.92)&0.61(0.67) \\
CDF&$\sqrt{s}=1.8$ TeV&17.8$^\star$(19.9)&0.94(0.97)&0.65(0.67) \\
ISAJET&40 TeV, $\vev{p_J}=73\,\hbox{GeV}/c$&26.2(25.2)&
1.04(1.00)&0.69(0.67) \\
ISAJET&40 TeV, $\vev{p_J}=677\,\hbox{GeV}/c$&69.8(64.7)&1.15(1.06)
&0.72(0.67)  \\
\hline
\hline
\multicolumn{5}{l}{$^\star$ The extrapolated low-momentum part of the
function contributes 10 to this total.}\\
\end{tabular}
\end{center}
\end{table}

\begin{figure}
\centerline{\includegraphics{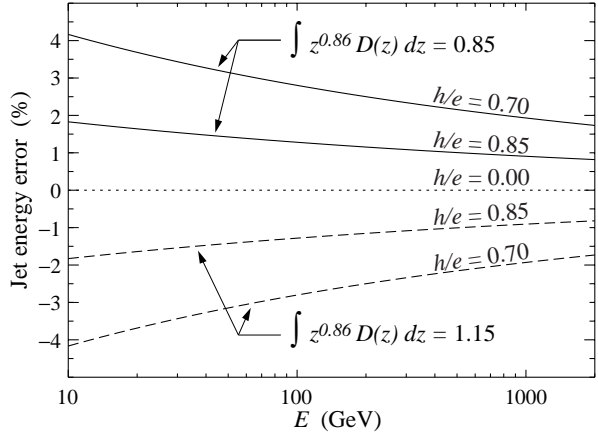}}
\caption{Energy determination error as a function of jet energy for 
representative values of $h/e$, for the two extreme case of the correction 
factor: 0.85 (top curves, for hard, low-multiplicity jets) and 1.15 
(bottom curves, for soft, high-multiplicity jets).
}
\label{fig:jeterror}
\end{figure}

The compensation factor $a\approx(1-h/e)$ appearing in 
Eq.~(\ref{eqn:jetsummation}) and~\ref{eqn:jetintegral} serves to further 
reduce the effect of the correction factor in producing a $jet/\pi$ 
difference. The percentage errors for the two limiting cases 0.85 and 1.15 
are plotted in Fig.~\ref{fig:jeterror} for calorimeters with $h/e = 0.70$ 
and $h/e = 0.85$, values which might occur for a Pb/LAr or badly designed 
metal/scintillator calorimeter.  The uncertainty in the exponent $m$ could 
introduce an error of about 3\% for jets below 100~GeV in a poor 
calorimeter.

In the context of a power law approximation to the hadronic fraction for 
an incident pion, the response for an incident jet thus differs
from the response to an incident pion by a simple correction 
factor, an integral over the fragmentation function. Given the 
uncertainties involved, no difference between jet and pion response can be 
found.

\section{Beating the devil}\label{sec:devil}

An estimation of the $\piz$ content of individual events would permit 
correction for intrinsic fluctuations (reduction of the ``constant 
term''), along with its contribution to energy uncertainty.

Several attempts have been made to use the radial and longitudinal detail 
to estimate, and correct for, the $\piz$-induced cascades.  During tests for the SDC 
construction, it was proposed that the $\piz$ contribution might come 
``early'' in the cascade, and could be estimated by excess energy deposit 
in the first layers.  This turned out not to be true\cite{dangreen}.  The 
ATLAS barrel calorimeter group adjusts downward the contribution of 
readout cells with large signals, since these tended to be from $\piz$
cascades.  In the test-beam runs they achieved slight improvements, 
e.g., from
$(46.9\pm1.2)\%/\sqrt{E}$ to $(45.2\pm2.2)\%/\sqrt{E}$\cite{ATLAScavalli}. 
Nearly a decade earlier, this algorithm was successful in correcting the WA~78
data\cite{wa78}.
Ferrari and Sala have simulated such 
corrections for the LAr TPC ICARUS detector, where very detailed 3D 
imaging is possible, and conclude that the $\piz$ content can be fairly well 
determined from the direct observation of the em cascades\cite{ferrari00}. 
The success of these corrections depends on the detail available, and any 
gains are usually marginal.

Mockett\cite{historyDREAM} suggested long ago that information from a 
dual-readout calorimeter 
with different $h/e$'s in the two channels could be used to estimate the 
electromagnetic fraction $f_{\pi^0}$ for each event.  Winn\cite{winn89} has 
proposed using ``orange" scintillator, observing the ionization 
contribution through an orange filter and observing the Cherenkov 
contribution through a blue filter. This has not yet been implemented, and 
looks problematical.

The idea of using a quartz fiber/scintillator fiber dual readout 
calorimeter to extract an estimate of $E_\piz$ for 
each event was discussed by Wigmans in 1997\cite{protoDREAM97}.
Since then,
the DREAM collaboration(Akchurin et~al.\cite{DREAM05}) has elegantly 
demonstrated the efficacy of the dual-readout technique, using a 
copper/optical fiber test-beam calorimeter. It consists of copper tubes, 
each containing three plastic scintillator fibers and four undoped fibers 
which produce only Cherenkov light.  These are read out separately for 
each event.

\begin{figure}
\centerline{\includegraphics{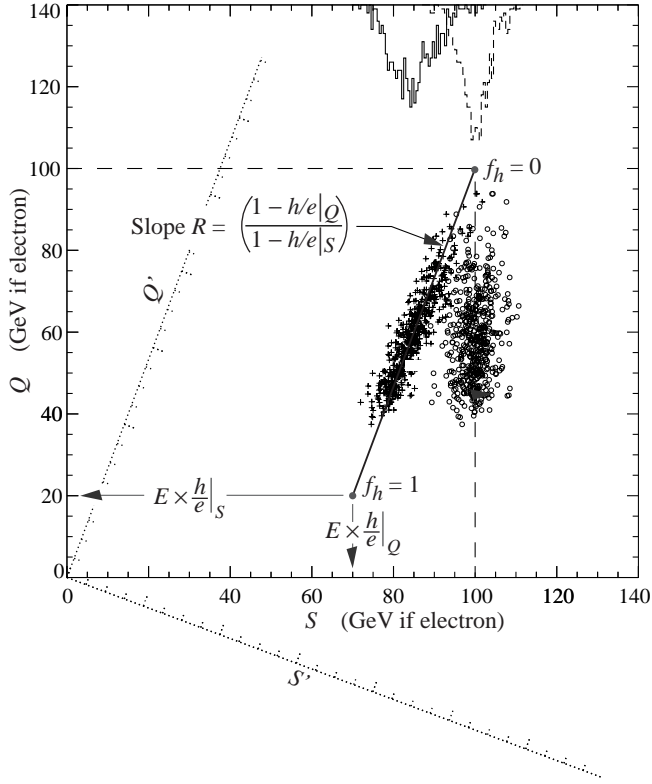}}
\caption{A toy model showing energy correction for 100 GeV pions in an 
idealized DREAM calorimeter, where $Q$ is the response in the quartz-fiber 
readout and $S$ is the response in the scintillator-fiber readout.  The 
observed ``events'' are shown by the {\scriptsize+}'s, and the corrected 
events by the $\circ$'s.  Rotating to a frame in which the $Q^\prime$ axis is 
parallel to the event locus provides an equivalent reduction.
}
\label{fig:q_vs_s}
\end{figure}

The principle is illustrated in Fig.~\ref{fig:q_vs_s}.  Akchurin et~al.'s 
notation is used: $S$ for the scintillator signal and $Q$ for the 
Cherenkov signal, with both energy scales calibrated with electrons. For 
this example their values $h/e|_S=0.7$ and $h/e|_Q = 0.2$ are used 
for the scintillator and nominal Cherenkov readouts respectively.  If there were 
only resolution contributions from the $E_\piz$ distribution,
events with different $f_{\pi^0}$ should lie along 
the solid line drawn from $f_h =1$ to $f_h =0$ ($f_{\pi^0} =0$ to $f_{\pi^0} =1)$:
\begin{equation}
S = E(f_{\pi^0}+ f_h(h/e|_S))\ , \hbox{\ and} \ \
Q = E(f_{\pi^0}+ f_h(h/e|_Q))\ .
\label{eqn:SQdefs}
\end{equation}

The effects of finite resolution are illustrated using simulations 
which give $f_{\pi^0}$ for 100~GeV negative pions axially incident on a 
very large lead cylinder.  (Results at 30~GeV from the same study are 
shown in Fig.~\ref{fig:f0_distr}(a).) 
For this cartoon example I arbitrarily 
introduced a Gaussian scatter in both $E_{\pi^0}$ 
($\sigma_{\pi^0}/\sqrt{100~\hbox{GeV}} = 1.5\%$) 
and $E_h$ ($\sigma_h/\sqrt{100~\hbox{GeV}}  = 3.0\%$).
The ``events'' are shown by the small +'s 
in the figure.  The solid histogram at the top shows the marginal 
distribution in $S$.  The mean is 84.7~GeV, the fractional standard deviation is 
5.3\%, and there is the usual skewness toward high energies.

Energy correction is straightforward. With the definition
\begin{equation}
R = \frac{1-h/e|_Q}{1-h/e|_S}\ ,
\label{eqn:Rdef}
\end{equation} 
Eq.~(\ref{eqn:SQdefs}) can be solved to obtain
\begin{equation}
E_{\rm corr} = \frac{RS - Q}{R-1}  \ .
\label{eqn:Ecorrected}
\end{equation}
The circles in Fig.~\ref{fig:q_vs_s} show the same events after reduction 
via Eq.~(\ref{eqn:Ecorrected}), and the dashed histogram shows the marginal 
distribution. The mean is 100.1~GeV, the fractional standard deviation is 3.4\%, and 
there is no evident skewness.  Complete compensation has been achieved 
by using the simultaneous  readouts.

Alternatively, a coordinate rotation to axes ($Q^\prime,S^\prime$) can be 
made, such that the new $Q^\prime$ axis is parallel to the event locus. 
The projection of the event distribution onto the new $S^\prime$ axis is 
of minimal width. Scaled upward by the geometrical factor, it becomes the 
corrected distribution given by Eq.~(\ref{eqn:Ecorrected}).

The importance of a large ``compensation asymmetry'' is evident.  If the 
standard deviation in $Q$ is $\sigma_Q$ and the standard deviation in $S$ 
is $\sigma_S $ (both in GeV), then the variance in $E_{\rm corr} $ is 
approximately
\begin{equation}
\sigma_{\rm Ecorr}^2 = \left(\frac{R}{R-1}\right)^2\sigma_S^2  + 
\left(\frac{1}{R-1}\right)^2\sigma_Q^2
\label{eqn:ecorr_error}
\end{equation}

Since $R$ can be well determined either from test-beam measurements of
$\pi/e$ as a function of energy or from fits to the slope in a plot of $Q$
vs $S$ at one energy, the error in $R$ has been neglected in writing
Eq.~(\ref{eqn:ecorr_error}).  In the present example $R = (1-0.2)(1-0.7) =
2.66$, so $\sigma_E^2= 3.20\,\sigma_S^2 + 0.36\,\sigma_Q^2$.  Given the
Cherenkov readout, $\sigma_Q^2$ is likely to be much larger than
$\sigma_S^2$.  The price of the correction is an increased error on each
event, but it is clear from Fig.~\ref{fig:q_vs_s} that there is
compensating improvement.

Alternate schemes simpler than DREAM would be desirable. Winn's
scheme\cite{winn89}, taking advantage of the different colors of Cherenkov
and (red) scintillation light, uses common detectors but still needs the
doubled number of photomultipliers. Clean separation of the two signals  
would likely be difficult.  LSND\cite{LSND} used a weak scintillator and
distinguished between directional Cherenkov light and isotropic
scintillation light. But this was a very different kind of detector, a  
homogeneous low-$Z$ detector used in a search for rare signals involving
single, low-energy electrons.  The electrons at the end of a high-energy
shower do not remember the original direction.

Descendants of DREAM  are being studied under the rubric of the 
``Fourth Concept Detector'' for the International Linear
Collider\cite{ILC4th}.  One starting point might be the dual-readout quartz/scintillator
DREAM concept.  It can immediately be improved to get rid of signal correlation between
adjacent fibers and simplified, e.g., by alternating 
scintillator and Cherenkov fibers in groves in sheets of absorber 
(possibly tungsten) which can then be combined as a sandwich.  

An interesting departure from the fiber calorimeter idea 
was presented by Zhao at a March, 2006 ILC
workshop\cite{zhao06}.  A ``conventional'' iron/scintillator-plate
sandwich calorimeter is constructed in which lead glass tiles are
substituted for all or part of the iron plates.  The Cherenkov light is
detected via waveshifter fibers in grooves in the tiles.  Since lead glass
(heavy flint glass) with specific gravities up to about 5.7 are available,
the calorimeter thickness might not be that excessive.  However, one
expects that transverse structure observation is essential, and 
there remains the problem of neutron detection.

Perhaps a better approach is to observe
the fast, blue, directional, polarized Cherenkov light from an inorganic
scintillator.  This is the object of present test beam work, using the scintillator 
PbWO$_4$\cite{PbWO4}.  (The slow
component has only a 50~ns decay time,  $\lambda_{\rm max}=560$~nm (yellow), 
but the scintillation efficiency is only 0.1\% that of~NaI.)   

But if neutron detection can be added to neo-DREAM, then the 
in-principle ultimate 1\% hadron calorimeter resolution\cite{protoDREAM97}
might be approached.  Ways to do this are under active investigation by the 
Forth Concept collaborators and others, who hope to exploit one or more
distinguishing features of the neutron signal:

\begin{enumerate}

\item
Neutrons distribute further from the core of a cascade than do other 
components. 

\item
Gamma rays from nuclear de-excitation following thermal neutron capture
are slow, in the several hundred ns range\cite{wigmansbook}. 
Given fast gating requirements, their signal is probably not useful.

\item
In a hydrogenous scintillator, ionization from the proton recoils in $n$-$p$ elastic scattering
can be observed.

\item 
Most neutron detectors in nuclear physics take advantage of the large
cross sections for the $^{10}{\rm B}(n,\alpha)$, $^6{\rm Li}(n,\alpha)$,
and $^3{\rm He}(n,p)$ reactions, with the boron reaction being the most
popular\cite{knoll00}. The gas-filled detectors in common use are
impractical for a calorimetric applications, but boron- and lithium-loaded
scintillators exist and are being further developed. The inorganic scintillator
LiI(Eu) is an obvious candidate, but its crystalline structure and 300~ns
decay time both present problems.  Organic borate additives in conventional
plastic scintillators might have promise.  There are common high-boron 
glasses and there are glass scintillators; whether a boron glass can be
made to scintillate remains to be seen.  But in any case, the time scale involved
here is probably much too long.

\item 
Neutron interaction products are slow protons and fission fragments, so the nonlinear 
light output in scintillators (Birks' law) offers another, if unlikely, avenue.

\end{enumerate}

One might imagine planes of PbWO$_4$ functioning as a dual readout calorimeter, 
with interleaved organic scintillator sheets.  The  PbWO$_4$ produces scintillation light
from several ionization processes; the hydrogenous scintillator does the same (weighted
a little differently), but it also detects the ionization from the $n$-$p$ elastically
scattered protons. 
For each event the PbWO$_4$ scintillation signal, the PbWO$_4$ Cherenkov
signal, and the organic scintillator signal might be 
represented as a point in a data cube 
analogous to the two-dimensional ``data square'' shown in 
Fig.~\ref{fig:q_vs_s}.  It will be interesting
to see if a correction formula as simple as Eq.~\ref{eqn:Ecorrected} 
can be found.

With sufficient work and a little luck, it seems likely that the Holy Grail of ultimate
resolution will be implemented in future calorimeters.

\section{Discussion}

The conceptual picture of the physics of a hadronic cascade and the 
scaling law it implies have been rich in consequences for understanding 
the behavior of a hadron calorimeter.  The response ratio $\pi/e$ is particularly simple, 
and the pion-proton 
response difference, in retrospect so obvious, was an unexpected surprise. 
If the incident hadron is a jet rather than a pion, the response is still 
given by Eq.~(\ref{eqn:meanvis}), except that $f_h$ is multiplied by an 
integral over the fragmentation function which appears to be near unity. 
The recent results using dual readout are explored and possible extensions are discussed.
Resolution is described by considering in detail the various stochastic processes 
involved in a hadronic cascade; they cannot be simply convoluted.

\section*{Acknowledgments}
I  am indebted to a large fraction of my calorimetry and radiation physics
friends for profitable discussions in the course of formulating these
ideas, but particularly so to Richard Wigmans and my collaborators on
Paper~I:  Tony Gabriel, P.K. Job, Nikolai Mokhov, and Graham
Stevenson.  Conversations with and input from Nural Akchurin, 
Alberto Fass\'o, Alfredo Farrari,
and John Hauptman have been especially welcome and useful.  
Sec.~\ref{sec:mips} was written in close collaboration with Hans Bichsel.
Ed Wang made the ISAJET simulations used in Sec.~\ref{sec:jetresponse}.

This work was supported by the U. S. Department of Energy under Contract
No.\ DE-AC02-05CH11231.

\appendix
\section{Resolution}\label{sec:resolution_appendix}

In the cascade initiated by a hadron with energy $E$, energy is 
transferred to the em sector via $\pi^0$ production and decay.  
The energy deposit in the resulting em cascades produce ionization
with some efficiency $e$. Most of the ionization is via energy loss 
by the abundant low-energy electrons.  
In the case of the CMS developmental Cu/quartz-fiber  test 
calorimeter\cite{qcal97}, Cherenkov light samples part of the electron path length.
Quite independently, the hadronic component produces ionization through
the many mechanisms involved in hadronic cascades, again mostly by ionization 
by low-energy particles, with overall efficiency $h$.
Each  goes its stochastic way independently
of the other. One must calculate the distribution of the sum of the contributions to the ionization with the constraint that  $f_\piz $ is fixed, then integrate over $f_\piz $.
Finally, the distribution 
is ``sampled'' by directly collection ions or detecting
scintillation light (or Cherenkov light) via photomultipliers or photodiodes. 
The resulting p.d.f. has a variance somewhat different than the usual 
$\sigma/E = \sigma^{\prime}/\sqrt{E} \oplus \sigma_{\rm const}$. 
The differences, discussed in Sec.~\ref{sec:resolution}, are easily 
understood physically.

To expedite the calculations, it is useful to associate a characteristic 
function (c.f.) $\vev{\e^{iux}} = \phi(u)$  with each p.d.f. $f(x)$.  
It is essentially the Fourier transform of the p.d.f., and
is discussed in the {\it Probability\/} section of RPP06\cite{RPP06} and 
many other places\cite{cramer}.  Among the properties I will use are:

\begin{itemize}

\item
Convolution of  p.d.f.'s becomes multiplication of c.f.'s:
\begin{equation}
f(x) = \int f_1(x) f_2(x-y) dy \quad\Longrightarrow\quad \phi(u) 
= \phi_1(u) \phi_2(u)
\label{eqn:convolve}
\end{equation}

\item
Let the conditional p.d.f.\  of $f_2(x|z)$ be $\phi_2(u|z)$ and 
the p.d.f.\ of $z$ be $f_1(z)$. Then
\begin{equation}
\phi(u) = \int f_1(z) \phi_2(u | z)dz \ .
\label{eqn:conditionall}
\end{equation}

\item
If  $\phi_2$ (above) is of the form $\phi_2(u|z) = 
A(u)\exp(ig(u)z)$, then
\begin{equation}
\phi(u) = A(u) \phi_1(g(u)) \ .\label{eqn:g_of_u}
\label{eqn:cf_of_Ag}
\end{equation}
where $\phi_1(u)$ is the c.f.\ of $f_1(z)$.

\item
The c.f.\ of a Gaussian p.d.f.\ with mean $m$ and variance $\sigma^2$ is 
\begin{equation}\phi(u) =
\exp{(imu - \sigma^2 u^2/2)} \ . 
\label{eqn:gaussiancf}
\end{equation}

\item
Higher moments may be included by continuing the series:
\begin{equation}
\phi(u) = \exp{(imu - \sigma^2 u^2/2 -i\mu_3u^3/3! )+ \ldots}
\label{eqn:skewedcf}
\end{equation}
Here $\mu_3$ is the third moment of the distribution about the mean. 
The dimensionless ``coefficient of skewness'' $\gamma_1 = \mu_3/\sigma^3$ 
was introduced in Sec.~\ref{sec:albedo} and will be used here.
\end{itemize}

The arrows between boxes in Fig.~(\ref{fig:EFlowSimple}) actually
indicate the various p.d.f.'s describing fluctuations in
each of the steps. A more complete version, Fig.~\ref{fig:EFlow_all}, defines
these distributions, which, along with their c.f.'s, means, and 
variances are given in Table~\ref{tab:resfunctions}.  
The notation is somewhat verbose in the interest of clarity.
Throughout the calculations, the primary hadron energy $E$ is 
implicit and constant.  Whether energies (such as $E_\piz$) or energies scaled by the incident
energy (such as $f_\piz = E_\piz/E$) are used as variables is arbitrary.  
I make the split choice of using $f_\piz$ and $f_h$,
and energies elsewhere, to prevent even more complex notation.

The p.d.f.\ $\Pi(f_\piz )$ is discussed in  Sec.~\ref{sec:albedo}. 
For reasons discussed there, 
$f_\piz$ is chosen as the independent variable rather
than its hadronic counterpart  $f_h = 1-f_\piz$.  Typical simulations 
are shown in Figs.~\ref{fig:comp30_bothdot} and \ref{fig:f0_distr}.  
The mean of $f_\piz$ was defined as $f_\piz^0$, 
the fractional variance was found to be $\sigma_\piz^2$,
and its coefficient of skewness $\gamma_1$ was found to be about 0.6. Its c.f.\  is thus
\begin{equation}
 \phi_\Pi(u) = \exp{(i u f_\piz^0 - u^2\sigma^2_\piz/2  - iu^3 \gamma_1\sigma_\piz^3 3!  + \dots)}\ .
\label{eqn:phiPi}
\end{equation}
The skewness is carried forward in the calculation. The other p.d.f.'s are 
assumed to be
near-Gaussian, with c.f.'s of the form given in Eq.~(\ref{eqn:gaussiancf}).

The conditional p.d.f.\ $g_\piz(E^{\rm vis}_\piz|f_\piz )$ describes the 
visible signal produced by the deposit of $E \,f_\piz $ in the em sector.  
``Visible'' means energy deposit, usually ionization,
which can be sampled by an appropriate transducer.
Its mean value is $eE_\piz $,\footnote{The normalization of $e$ and $h$ 
is ignored because at the end only the ratio $h/e$ appears.} 
and  its c.f.\ is  $\phi_{g\piz}(u|E_\piz )$.  The variance for
an ensemble of events with the same $E_\piz $ should be proportional to $E_\piz $.
The c.f. may be written as
\begin{equation}
\phi_{g\piz}(u| E_\piz ) =\exp(iu e  E_\piz  - u^2\sigma_{e0}^2 e E_\piz /2) \ .
\label{eqn:phi_ge}
\end{equation}
where $\sigma_{e0}^2$ scales the variance.  Since the variance has units of (energy)$^2$ and
is proportional to the energy, $\sigma_{e0}^2$ has the units of energy.

\begin{table}
\caption{Probability distribution functions (p.d.f.'s) and characteristic 
functions (c.f.'s)  used in the resolution discussion.  
The primary energy $E$ is an implicit conditional variable,
and the em energy $E_\piz $ is the independent conditional variable 
used in the development.
The p.d.f. of the final sampled energy is not used explicitly.}
\centerline{\vbox{
\halign{\hbox to1.9in {\ \ #\hss} &
              \hbox to 1.2in {\hss$#$} &
              \hbox to 1.0in {\hss$#$} &
              \hbox to 1.0in {\hss$#$} &
              \hbox to 1.0in {\hss$#$} \cr
\noalign{\vskip3pt\hrule\vskip3pt\hrule\smallskip}
     Distribution & \hbox{p.d.f.} & \hbox{c.f.} & \hbox{Mean} & \hbox{Variance}  \cr
\noalign{\smallskip\hrule\smallskip}
Fractional energy of $\pi^0$'s & \Pi(f_\piz ) & \phi_\Pi(u) & f_\piz  & \sigma_\piz ^2\cr
Ionization in em showers & g_\piz(E^{\rm vis}_\piz |E_\piz ) &\phi_{g\piz}(u|E_\piz ) 
      & eE_\piz  & eE_\piz  \sigma^2_{e0} \cr
Ionization by hadrons  &g_h(E^{\rm vis}_h |E_\piz ) & \phi_{gh}(u|E_\piz )
      & h(E-E_\piz ) & h(E-E_\piz ) \sigma^2_{h0} \cr
Total ionization, fixed $E_\piz $ & F_{\rm vis}(E^{\rm vis}|E_\piz )  
      & \phi_{\rm vis}(u | E_\piz ) & 
\hbox{Eq.~(\ref{eqn:phi_ion})} & \hbox{Eq.~(\ref{eqn:phi_ion})}\cr
Total ionization & F_{\rm vis}(E^{\rm vis})  & \phi_{\rm vis}(u) 
   & \hbox{Eq.~(\ref{eqn:finalphivis})} & \hbox{Eq.~(\ref{eqn:finalphivis})} \cr
Sampled signal, fixed $E^{\rm vis}$  & F_{\rm samp}(E^{\rm samp}|E^{\rm vis}) 
    & \phi_{\rm samp}(u|E^{\rm vis}) &  E^{\rm samp}-E^{\rm vis} 
    & eE^{\rm vis}\sigma_{\rm samp0}^2\cr
Final sampled signal &[ F_{\rm samp}(E^{\rm samp}) ]
     & \phi_{\rm samp}(u)   & 
    \hbox{Eq.~(\ref{eqn:piovere_again})} &  \hbox{Eq.~(\ref{eqn:resolution})}\cr
\noalign{\vskip3pt\hrule\vskip3pt\hrule\smallskip}
}}}
\label{tab:resfunctions}
\end{table}

Similarly, the distribution of ionizing hadronic energy at 
fixed $E_\piz $ is given by $g_h(E_h^{\rm vis}|E_\piz )$.  The mean is $h(E-E_\piz )$.
(Since $E_h = E-E_\piz $, it is sufficient to express the condition on 
$E_h$ as a condition on  $E_\piz $.)  In analogy to Eq.~(\ref{eqn:phi_ge}), 
I write the c.f. as 
\begin{equation}
\phi_{gh}(u | E_\piz ) =\exp(iu h (E-E_\piz ) - u^2\sigma_{h0}^2 h (E-E_\piz )/2 ) \ .
\label{eqn:phi_gh}
\end{equation}
The complicated hadronic response is dominated by a small number of 
collisions with 
large nuclear binding energy losses, so its distribution is wider than the em
response\cite{D0,akesson87,zeus_FCAL}.
It is thus expected that 
$\sigma_{h0}^2>\sigma_{e0}^2$, but as shown in Sec.~\ref{sec:resolution}  it is 
hard to distinguish the ways the contributions of $\sigma_{e0}^2$ and the 
sampling term modify the energy dependence of the resolution.

I interpret $g_\piz(E^{\rm vis}_\piz |E_\piz )$ and $g_h(E^{\rm vis}_h |E_\piz )$ 
as the $\piz$ and hadronic contributions, respectively, to the intrinsic 
resolution. This point will be explored later.

Only the total ionization (or Cherenkov light)  
$E^{\rm vis} = E_\piz ^{\rm vis}+E_h^{\rm vis}$ can be sampled.
Let the conditional  p.d.f.\ of $E^{\rm vis}$ be $F_{\rm vis}(E^{\rm vis}|E_\piz )$:
\begin{equation}
F_{\rm vis}(E^{\rm vis}|E_\piz )= 
\int g_\piz(E_\piz ^{\rm vis}|E_\piz ) \,g_h(E^{\rm vis}-E_\piz ^{\rm vis}|E_\piz ) dE_\piz ^{\rm vis}
\label{eqn:condfvis}
\end{equation}
This integral is a simple convolution, so by Eq.~(\ref{eqn:convolve})
\begin{equation}
\phi_{\rm vis}(u|E_\piz ) = \phi_{g\piz}(u|E_\piz ) \phi_{gh}(u|E_\piz ) \ .
\label{eqn:condphivis}
\end{equation}

The sum over $E_\piz $ results in the distribution
\begin{equation}
F_{\rm vis}(E^{\rm vis}) = 
   \int \Pi(f_\piz ) \, F_{\rm vis}(E^{\rm vis}|E_\piz ) dE_\piz  \ .
\label{eqn:finalfvis}
\end{equation}
Via Eq.~(\ref{eqn:conditionall}) the c.f.\ of $F_{\rm vis}$ is
\begin{equation}
\phi_{\rm vis}(u) = \int \Pi(f_\piz ) \, \phi_{\rm vis}(u|E_\piz ) dE_\piz  \ .
\end{equation}

The c.f.\ $\phi_{\rm vis}(u|E_\piz )$ can be calculated using 
Eqs.~(\ref{eqn:phi_ge}) and~(\ref{eqn:phi_gh}).  For simplicity here 
and in the algebra 
leading to Eq.~(\ref{eqn:finalphivis}), it is convenient to define 
$\Delta\sigma^2 =  \sigma^2_{h0} h/e-\sigma^2_{e0}$. 
Terms involving $E_\piz $ are collected into the second exponential:
\begin{equation}
\phi_{\rm vis}(u | E_\piz ) = 
e^{iuhE  - u^2\sigma^2_{h0}  hE/2}\times e^{i e E_\piz ( u (1-h/e) - iu^2\Delta\sigma^2/2)} 
\label{eqn:phi_ion}
\end{equation}

Written in this way, $\phi_{\rm vis}(u|E_\piz )$ is of the 
form $A(u) \exp(ig(u) z)$, so by Eq.~(\ref{eqn:cf_of_Ag}),
\begin{equation}
\phi_{\rm vis}(u) = 
       e^{iuhE  - u^2\sigma^2_{h0}  hE/2} \times \phi_\Pi(u e (1-h/e)- 
       iu^2 e \Delta\sigma^2/2) \ ,
\label{eqn:phivis}
\end{equation}
where $g(u)$ is identified with $ e ( u (1-h/e) - iu^2\Delta\sigma^2/2 )$.
$\phi_\Pi(u)$ is given by Eq.~(\ref{eqn:phiPi}), so it remains 
to substitute this function into Eq.~(\ref{eqn:phivis}) and collect
the terms multiplying powers of $u$.  These terms can then be identified 
as the mean, variance, and skewness of $F_{\rm vis}(E^{\rm vis})$.  
After considerable algebra, Eq.~(\ref{eqn:phivis}) yields
\begin{eqnarray}
\phi_{\rm vis}(u) =& \exp\Big(iueE(f_\piz^0 + f_h^0h/e)      \Big.
\nonumber\\
-& \textstyle{\frac{1}{2}}u^2 eE[f_\piz\sigma_{e0}^2 + f_h^0 \sigma_{h0}^2 h/e + 
(1-h/e)^2  \sigma_\piz^2 e ^2 E^2] 
\nonumber\\
-& \Big.\textstyle{\frac{1}{3!}}u^3 [\gamma_1 \sigma_\piz^3 e^3 E^3(1-h/e)^3 +
      3\sigma_\piz^2 \Delta\sigma^2 e^2E^2(1-h/e) \Bigr] + \ldots \Big)
      \phantom{mmm}  
\label{eqn:finalphivis}
\end{eqnarray}

The final step is to ``sample'' the ionization with whatever output 
transducer is being used.  Although the experimenter has little control 
over the variance of $F_{\rm vis}(E^{\rm vis})$.%
\footnote{There are two caveats here: The effects of noncompensation 
can be minimized by the methods used by the DREAM 
collaboration\cite{DREAM05}, as discussed in Sec.~\ref{sec:devil}
and (in principle so far) by measuring the neutron flux on an 
event-by-event basis\cite{wigmans98_neut}
in order to reduce the intrinsic resolution contribution of 
$g_h(E_h^{\rm vis}|E_\piz )$.}
the design might be changed to improve light collection, for example,  
if the variance contribution due to photoelectron
statistics were significant.  Again a Gaussian distribution is assumed. 
The variance contribution from the sampling transducer is proportional 
to $E^{\rm vis}$:
\begin{eqnarray}
F_{\rm samp}(E^{\rm samp}|E^{\rm vis}) 
    =& \frac{1}{\sqrt{2\pi  \sigma_{\rm samp0}^2 E^{\rm vis}}}
    \exp\left[-\frac{(E^{\rm samp}-E^{\rm vis})^2}
    {2 \sigma_{\rm samp0}^2 E^{\rm vis}}\right]\\
\phi_{\rm samp}(u|E^{\rm vis}) 
    =& \exp[iE^{\rm vis}(u + \textstyle{\frac{i}{2} }u^2\sigma_{\rm samp0}^2)]\\
\phi_{\rm samp}(u) 
=& \int F_{\rm vis}(E^{\rm vis}|E_\piz )\phi_{\rm samp}(u|E^{\rm vis})dE^{\rm vis}
\end{eqnarray}
Since the variance is not a constant, a simple convolution is again 
insufficient.  Following the recipe of Eq.~(\ref{eqn:cf_of_Ag}), $g(u) = 
u + iu^2\sigma_{\rm samp0}^2/2$ is substituted for $u$ in 
Eq.~(\ref{eqn:finalphivis}):
\begin{equation}
\phi_{\rm samp}(u) = \phi_{\rm vis}(u +
 iu^2\sigma_{\rm samp0}^2/2)
\label{eqn:finalphisamp}
\end{equation}

The mean pion response (the multiplier of $iu$ 
in Eq.~(\ref{eqn:finalphivis}) is unaffected:
\begin{eqnarray}
\hbox{``}\pi\hbox{''} = &eE (f_\piz^0 + f_h^0h/e)  \ ; \nonumber\\
                       {\rm or}\quad  \pi/e     = & 1 - (1-h/e)f_h^0 \ ,  
\label{eqn:piovere_again}
\end{eqnarray}
so that the usual form for $\pi/e$ (Eq.~(\ref{eqn:meanvis}) is recovered.

However, $eE (f_\piz^0 + f_h^0h/e) \sigma_{\rm samp0}^2$ is added
to the variance of $F_{\rm vis}(E^{\rm vis})$  
(the multiplier of $-iu^2/2$ in Eq.~(\ref{eqn:finalphivis})).  
The final fractional variance for the calorimeter is
\begin{eqnarray}
\left({\sigma\over E}\right)^2 =&
\frac{(f_\piz^0 + f_h^0h/e) \sigma_{\rm samp0}^2 }{E} +
\left[\frac{f_\piz^0\sigma_{e0}^2 }{E} + \frac{f_h^0 \sigma_{h0}^2 h/e}{E} 
\right] + (1-h/e)^2\sigma_\piz^2 \nonumber \\
=& \frac{(\pi/e) \sigma_{\rm samp0}^2 }{E}
+ \left[ \frac{(1-f_h^0)\sigma_{e0}^2 }{E}+ \frac{f_h^0 \sigma_{h0}^2 h/e}{E}
\right] + (1-h/e)^2\sigma_\piz^2(E) \ ,\phantom{mmmm} \label{eqn:resolution}
\end{eqnarray}
where I have followed convention and scaled the energy to electron calibration: 
$eE  \to E$. The energy dependence of $\sigma^2_\piz$ is made explicit in the last line.
This is {\it almost} the usual form for the resolution 
($\sigma/E = C_1/\sqrt{E} \oplus C_2$), but with some important differences:

\begin{enumerate}

\item
The first term, the sampling contribution, is scaled by the em response.  
Since it is the ionization
which is sampled, this contribution is perforce proportional to 
$\pi/e$.

\item The terms in square brackets are the em and hadronic contributions 
to the intrinsic variance.  The shape and interpretation 
of these terms is discussed in Sec.~\ref{sec:resolution}.

\item The analysis reproduces the familiar ``constant term,'' with variance 
contribution explicitly proportional to $(1-h/e)^2$.  Its important energy dependence is
discussed in Sec.~(\ref{sec:albedo}).

\end{enumerate}

The third moment about the mean ($\mu_3$) of the sampled distribution 
is the coefficient of $-iu^3/3!$ in $\phi_{\rm samp}(u)$:
\begin{equation}
\mu_3 = \gamma_1 \sigma_\piz^3  E^3(1-h/e)^3 +
      3\sigma_\piz^2 \Delta\sigma^2 E^2(1-h/e)
+ 3 E\sigma_{\rm vis}^2\sigma_{\rm samp0}^2 \ ,
\end{equation}
where the energy is again scaled to the electron calibration: $eE \to E$. Here
$E\sigma_{\rm vis}^2$ is the variance of $F_{\rm vis}(E_{\rm vis})$, the 
coefficient of $-u^2/2$ in Eq.~(\ref{eqn:finalphivis}).

The first term is to be expected in any noncompensating calorimeter; 
it is just the skewness of $\Pi(f_\piz )$ ``playing through'' to the end.  
As discussed in Sec.~\ref{sec:albedo}, the 
dimensionless coefficient of skewness, $\gamma_1$, is about 0.06 for the model
discussed there ($\pi^-$ on Pb, using an old version of FLUKA), and 
$\sigma_\piz = 12.5\%$ at 100~GeV with some
mild energy dependence.   $\gamma_1 \sigma_\piz^3 E^3$ is the 
actual third moment about the mean of $\Pi(f_\piz )$.

It is interesting that the visible energy deposition and sampling 
terms also contribute to the skewness.  In the first case, this is 
because the variance of the visible energy at fixed $E_\piz $ is proportional to 
$E_\piz $, and so at large $E_\piz $ a wider distribution is contributed to 
$F_{\rm vis}(E^{\rm vis})$ than for low $E_\piz $---even though 
for a given $E_\piz $ the distribution is (taken to be) Gaussian.

For the same reason, sampling also contributes to the skewness. 
The first two contributions both vanish if $h/e=1$, but the third 
term does not. Even in the case of a compensating calorimeter, we
should not expect an exactly Gaussian distribution.



\begin{thebibliography}{99} 

\bibitem{gabriel94}{T.A. Gabriel, D.E. Groom, P.K. Job, N.V. Mokhov,
and G.R. Stevenson, Nucl.\ Instr.\ and Meth.\ A 338 (1994) 336--347.}

\bibitem{tuscaloosa90}{D.E. Groom, Energy scaling of low-energy neutron yield,
the $e/\pi$ ratio, and hadronic response in a calorimeter, Proc.\
Workshop on Calorimetry for the Superconducting Super Collider, Tuscaloosa,
Alabama, 13--17 March 1989, eds., R. Donaldson and M.G.D. Gilchrise, World
Scientific, (1990) 59--75.}
 
 \bibitem{wigmansbook}R. Wigmans, {\em Calorimetry: Energy Measurement in Particle Physics, International Series of Monographs on Physics}, vol.~107, Oxford University Press (2000).

 \bibitem{barcelona89}D. E. Groom, Energy Scaling of Low-Energy Neutron Yield, the $e/\pi$ Ratio, 
 and Hadronic Response in a Calorimeter,  Proc.\ of the ECFA Study Week on Instrumentation 
Technology for High-Luminosity Hadron Colliders, Barcelona, Spain, 14--21 Sept. 1989, 
ed. by E. Fernandez and G. Jarlskog, CERN 89-10, 549--550, and ECFA, (1989) 89--124.
 
\bibitem{snowmass90}D. E. Groom, Contributions of Albedo and Noncompensation to Calorimeter 
Resolution,  Proc.\ of the 1990 DPF Summer Study on High Energy Physics Research 
Directions for the Decade, Snowmass CO, June 25--July 13, 1990, ed. by E. L. Berger and R. Craven, 
World Scientific, (1992) 403--406.
 
 \bibitem{fortworth90}D. E. Groom and E. M. Wang, Jet Response of a Homogenous Calorimeter, 
 Proc.\ of the Fort Worth Symposium on Detector R\&D for the SSC, Forth Worth TX, 15--18 Oct. 1990, 
 ed. by M. G. D. Gilchriese and V. Kelly, World Scientific (1991) 385--387.
  
 \bibitem{aachen90}
 D. E. Groom and E. M. Wang, Jet Response of an Ideal Calorimeter, Vol. III, 
 Proc.\ of the ECFA Large Hadron Collider Workshop, Aachen (4--9 October 1990), 
 CERN 90-10 (1990) 315--319.

\bibitem{capri91}D. E. Groom, Four-Component Approximation to Calorimeter Resolution, 
Proc.\ II Inter. Conf. on Calorimetry in High Energy Physics, Capri, Italy, 14--18 October 1991, 
ed. by A. Ereditato, World Scientific (1992) 376--381.
 
 \bibitem{tucson97}D. E. Groom, Energy flow in a hadronic cascase: Application to
 hadron calorimetry (invited talk), Proc.\ VII Inter.\ Conf.\ on Calorimetry
in High Energy  Physics, Tucson, Arizona, 9--14 November 1997, ed. E.
Cheu, T. Embry, J. Rutherfoord, R. Wigmans, World Scientific (1998) 507--521.


\bibitem{DPM} A. Capella, et al., Phys.\ Rep.\ 236 (1994) 225.

\bibitem{aliceFLUKA}{{\tt http://aliceinfo.cern.ch/static/Offline/fluka/manual/}}

\bibitem{EGS4}{W.R.~Nelson, H.~Hirayama, and D.W.O.~Rogers,
The EGS4 Code  System, SLAC-265, Stanford Linear Accelerator Center
(Dec.~1985)}.  FLUKA now contains its own em radiation transport code; this and other codes
have largely supplanted EGS in high-energy physics applications.

\bibitem{qcal97}N. Akchurin, et al., Nucl.\ Instr.\ and\ Meth.\ A 399 (1997) 202.

\bibitem{wigmans_pi_p_98}N. Akchurin, et al.,
Nucl.\ Instr.\ and\ Meth.\ A 408 (1998) 380.

\bibitem{wigmansAnnRev}R. Wigmans, Ann.\ Rev.\ Nucl. Part.\ Sci.\ 41 (1991) 133.

\bibitem{leroy00}C. Leroy and P-G. Rancoita, 
Rep.\ Prog.\ Phys.\ 63 (2000) 505.

\bibitem{ferrari00}A.~Ferrari and P.~R.~Sala, Physics processes in hadronic showers
(invited talk), Proc.\ IX Inter.\ Conf.\ on
Calorimetry in High Energy Phys., Annecy, October 9--14 2000, B.~Aubert,
J.~Colas, P.~N\'edelec, L.~Poggioli eds., Frascati Physics Series, (2001) 31--55.   

\bibitem{spacal92a}{D. Acosta, et al., Nucl.\ Instr.\ and\ Meth.\ A 316 (1992) 184.}

\bibitem{betafunction}C. Walck, Internal Rep.\ SUF-PFY/96-01, 
Fysikum, Univ.\ Stockholm (last modification 22 May 2001).

\bibitem{wa78}
M. De Vincenzi, et al., Nucl.\ Instr.\ and\ Meth.\ A~243 (1986) 348.

\bibitem{wigmans98_neut} R. W. Wigmans, Rev.\ Sci.\ Instr.\ 69 (1998) 3723.

\bibitem{eloss_to_muons}J. Alvarez-Mu\~niz, R. Engel, T. K. Gaisser, J. A. Ortiz, \& T. Stanev,
Phys.\ Rev.\  A~{\bf 69}, (2004) 103003.

\bibitem{behrens90}{J. Behrens, et al., Nucl.\ Instr.\ and\ Meth.\ A 289 (1990) 115.}

\bibitem{D0} S. Abachi, et al., Nucl.\ Instr.\ and\ Meth.\ A 324 (1993) 53--76.

 \bibitem{spacal91}{D. Acosta, et al., Nucl.\ Instr.\ and\ Meth.\ A 308 (1991) 481.}

\bibitem{jinbo98}J. B. Liu, Testbeam results for the CDF endplug
hadron calorimeter, Proc.\ VII Inter.\ Conf.\ on Calorimetry
in High Energy  Physics, Tucson, Arizona, 9--14 November 1997, ed. E.
Cheu, T. Embry, J. Rutherfoord, R. Wigmans, World Scientific, (1998) 237--240.

\bibitem{PbWO4}R. Wigmans, private communication (2006).

\bibitem{hangingfile}{A. Beretvas, et al., Nucl.\ Instr.\ and\ Meth.\ A 329  (1993) 50--61.}

\bibitem{fano63}U. Fano, Ann.\ Rev.\ Nucl.\ Sci.\ 13 (1963) 1.

\bibitem{rossi52}B. Rossi, {\em High-Energy Particles}, (Prentice-Hall, Inc., Englewood Cliffs, NJ, 1952).

\bibitem{ADNDT01}
D. E. Groom, N. V. Mokhov, and S. I. Striganov, Atomic and Nuclear Data Tables 78 (2001) 183.

\bibitem{RPP06}
W.-M. Yao, et al., The Review of Particle Physics, J. Phys.\ G  (2006) 1.

\bibitem{vavilov57}
P.~V. Vavilov, Sov.\ J. Phys.\ JETP 5 (1957) 749.

\bibitem{bichsel88}  H. Bichsel, Rev.\ Mod.\ Phys.\ 60 (1988) 663.

\bibitem{landau44}L. Landau, J. Phys.\ VIII (1944) 201; P.~V. Vavilov, Sov.\ J. Phys.\ JETP 5 (1957) 749.

\bibitem{talman79}R. Talman, Nucl.\ Instr.\ and\ Meth.\ 159 (1979) 189.

\bibitem{bichsel06pc}H. Bichsel, private communications (2006).

\bibitem{bichselAMOPH}H. Bichsel, Ch.~87 in the Atomic, Molecular
and Optical Physics Handbook, G. W. F. Drake, editor (Am.\ Inst.\ Phys.
Press, Woodbury NY, 1996).

\bibitem{bichsel06}H. Bichsel, 
Nuc.\ Instr.\ and Meth. A 562 (2006) 154--197. 

\bibitem{lohmann85}W. Lohmann, R. Kopp, and R. Voss,
Energy loss of muons in the energy range 1--10000 GeV,
CERN Report 85-03 (1985).

\bibitem{AtomicNuclearProperties}{\tt http://pdg.lbl.gov/AtomicNuclearProperties}

\bibitem{vanginneken86}A. Van Ginneken, Nucl.\ Instr.\ and\ Meth.\ A 251 (1986) 21.

\bibitem{striganov06}N. V. Mokhov, S I. Striganov, A. V. Uzunian, 
 On Fluctuations of Energy Losses of Ultrarelativistic Muons, (in Russian),     
 IFVE-80-56, (Serpukhov, IHEP) (Apr 1980), 12~pp.                             

\bibitem{akesson87}T. \AA kesson, et al.,
Nuc.\ Instr.\ and Meth.\  A 262 (1987) 243. 

\bibitem{spacal92}{D. Acosta, et al., Nucl.\ Instr.\ and\ Meth.\ A 320 (1992) 128.}

\bibitem{wigmans87}{R. Wigmans, Nucl.\ Instr.\ and\ Meth.\ A 259 (1987) 389.}  

\bibitem{bonnersphere}
A. Demianov, et al., CMS Internal Note CMS IN 2000/020 (February, 2000).

\bibitem{bonner60}
R. L. Bramblett, R. I Ewing, and T. W. Bonner, Nucl.\ Instr.\ and Meth.\  9 (1960)~1.

\bibitem{MARS96}
I. Azhgirey, et al., Nucl.\ Instr.\ and\ Meth.\ A 408 (1998) 535.

\bibitem{ILCdual}  R. Wigmans et al, ``Dual-Readout Calorimetry for High-Quality Energy
  Measurements'', October 2001, proposal to Advanced Detector
  Research Program of DoE; {\tt http://www.\-phys.ttu.\-edu/\-dream}

\bibitem{zeus_proton}A. Andreson, et al., 
Nucl.\ Instr.\ and\ Meth.\ A 336 (1993) 23.


\bibitem{drews90}{G. Drews, et al., 
Nucl.\ Instr.\ and Meth.\ A 290 (1990) 335.}  

\bibitem{delphi}{P. Aarnio \etal, Phys.\ Lett.\  240B  (1990) 271.} 

\bibitem{CDFjets}{F. Abe \etal, Phys.\ Rev.\ Lett.\ 65  (1990) 968.}  

\bibitem{ISAJET}{F.~E.~Paige and S.~D.~Protopopescu, Physics of the
Superconducting Supercollider, ed.\ by R. Donaldson and J.~Marx (Snowmass
CO, 1986),~320.}

\bibitem{dangreen} D. Green, private communication, about 1990.

\bibitem{ATLAScavalli}F. Ariztizabal, et al., Nucl.\ Instr.\ and\ Meth.\ A 349 (1994) 384.

\bibitem{historyDREAM} 
P. Mockett,  A review of the physics and technology of high-energy 
calorimeter devices, Proc.\ 11th SLAC Summer Inst.\ Part. Phy., July 1983, 
SLAC Report No. 267 (July 1983).

\bibitem{winn89}D. R. Winn and W. A. Worstell, 
IEEE Trans.\ Nuc.\ Sci 36, (1989)  334.

 \bibitem{protoDREAM97} R. Wigmans, Quartz Fibers and the Prospects
 for Hadron Calorimetry at the 1\% Resolution Level,
 Proc.\ VII Inter.\ Conf.\ on Calorimetry
in High Energy  Physics, Tucson, Arizona, 9--14 November 1997, ed. E.
Cheu, T. Embry, J. Rutherfoord, R. Wigmans, World Scientific, (1998) 182--193.

\bibitem{DREAM05}N. Akchurin, et al.,
Nucl.\ Instr.\ and\ Meth.\ A 537 (2005) 537

\bibitem{LSND} C. Athanassopoulos , et.~al., 
Nucl.\ Instr.\ and\ Meth.\ A 388, (1997) 149--172.

\bibitem{ILC4th} P. Le~Du, et al., ``Detector outline document for the Fourth Concept Detector
at the International Linear Collider,''\hfil\break
{\tt http://physics.uoregon.edu/$\sim$lc/wwstudy/concepts/\rm}
(May 2006).

\bibitem{zhao06}T. Zhao, ``Active Absorber Calorimeter,''  contribution 119 at 
Linear Collider Workshop 2006, Bangalore, India (8--13 March 2006);\hfil\break
{\tt indico.cern.ch/contributionDisplay.py?contribId=\hfil\break
119\&amp;sessionId=5\&amp;confId=568}

\bibitem{knoll00}G. F. Knoll, {\sl Radiation detection and measurement}, 
3rd edition, Wiley, New York (2000).

\bibitem{cramer}H. Cram\'er, Mathematical Methods of Statistics, Princeton
Univ.\ Press, New Jersey (1958).

\bibitem{zeus_FCAL}U. Behrens, et al., 
Nucl.\ Instr.\ and\ Meth.\ A 289 (1990) 115.

\end{thebibliography}
\end{document}